\documentstyle[preprint,aps,epsfig]{revtex}
\begin{document}
\title{Dilepton production in proton-nucleus and nucleus-nucleus
collisions at SPS energies}
\bigskip
\author{G. Q. Li and C. M. Ko}
\address{Cyclotron Institute and Physics Department, Texas A\&M University,\\
College Station, TX 77843, USA}
\author{G. E. Brown and H. Sorge}
\address{Department of Physics, State University of New York at Stony Brook,
Stony Brook, NY 11974, USA}
\maketitle
 
\begin{abstract}
Dilepton production in proton- and nucleus-induced reactions is studied 
in relativistic transport model using initial conditions determined by the
string dynamics from RQMD. It is found that both the CERES and HELIOS-3 
data for dilepton spectra in proton-nucleus reactions can be well described 
by the `conventional' mechanism of Dalitz decay and direct vector meson decay.
However, to provide a quantitative explanation of the observed dilepton 
spectra in central S+Au and S+W collisions requires contributions other 
than these direct decays.  Introducing a decrease of vector meson masses 
in hot and dense medium, we find that these heavy-ion data can also be 
satisfactorily explained.  This agrees with our earlier conclusions based 
on a fire cylinder model. We also give predictions for Pb+Au collisions 
at 160 GeV/nucleon using current CERES mass resolution and acceptance.
\end{abstract}
 
 
\section{Introduction}
 
In future experiments at the relativistic heavy-ion collider (RHIC) 
\cite{rhic}, dilepton production will play an important role in the 
search for signals of the quark-gluon plasma. Since dileptons interact 
only electromagnetically with the hadronic environment, they are thus 
penetrating probes of the early, most violent stage of heavy-ion 
collisions, where one expects that the quark-gluon plasma is likely 
to be created 
\cite{fein76,shur78,chin82,domo83,mcl85,kaja86,xia89,ko93,cley91,rus92,asak93}.
However, dileptons are produced in all stages of ultra-relativistic heavy-ion
collisions. In the initial stage, Drell-Yan processes are important and 
contribute essentially to dileptons with large invariant mass \cite{vogt}.
They are also produced from hadronic interactions as well as hadronic 
decays. These dileptons have most likely low to intermediate invariant 
masses. Thus, in order to deduce useful information about the quark-gluon 
plasma from dilepton spectra, it is important to have a good and accurate 
understanding of dilepton production from both initial Drell-Yan processes 
and later stage hadronic interactions.
 
Furthermore, the study of dilepton production from hadronic interactions 
carries useful information about hadron properties in hot and dense 
hadronic matter, which are expected to be different from those in 
free space due to the partial restoration of chiral symmetry.  There have 
already been suggestions to use dileptons to study the medium modification 
of pion dispersion relation at both finite density and temperature 
\cite{gale87,xia88}, the hadron electromagnetic form factor in dense 
matter \cite{wolf93}, and in-medium properties of vector mesons 
\cite{pis82,koch92,kar93,hat93,hoff94,li95a}. 
 
A particular useful aspect of dileptons is that their spectrum produced 
from the hadronic matter may reflect directly the in-medium vector meson 
masses, which have been shown in certain models to decrease with
increasing density and/or temperature
\cite{brown91,adami93,hatsu92,asa93,jean94,hatsu94,pwxia,thomas94}.
Dileptons have already been measured at Bevalac by the DLS collaboration
\cite{dls88} in heavy-ion collisions at incident energies around
1 GeV/nucleon. Theoretical studies have shown that the observed dileptons
with invariant mass above about 450 MeV are mainly from pion-pion
annihilation \cite{wolf93,xiong90}. Since the pion electromagnetic
form factor is dominated by the rho meson, measuring dileptons from
heavy-ion collisions thus provides the possibility of studying the 
properties of rho meson in hot and dense matter. Unfortunately, statistics 
are not good enough in the Bevalac experiments to give definite information 
on the in-medium rho meson properties. However, similar experiments with 
vastly improved statistics have been planned at SIS by the HADES 
collaboration \cite{hades94}.
 
On the other hand, recent observation of the enhancement of low-mass 
dileptons in central S+Au and S+W collisions at SPS/CERN energies by 
the CERES \cite{ceres95} and the HELIOS-3 \cite{helios95} collaboration, 
respectively, has generated a great deal of interest in heavy-ion community. 
Different dynamical models, such as hydrodynamical and transport models, 
have been used to investigate this phenomenon 
\cite{li95b,cass95a,cass95b,gale95}. Although several different mechanisms 
\cite{li95b,cass95a,cass95b,gale95,huang95,kapu95,wam95,koch96}
have been put forward to explain the observed enhancement, the most 
consistent explanation seems to be the decrease of vector meson masses 
in hot and dense matter \cite{li95b,cass95a,cass95b}.  This has been 
worked out by us using the relativistic transport model based on the 
extended Walecka model \cite{li95b}, and was supported by the calculation 
of Cassing {\it et al} \cite{cass95a,cass95b} using a dropping in-medium
rho meson mass predicted from the QCD sum rules.  As in hydrodynamical 
approaches \cite{gale95}, the starting point of Refs. \cite{li95b} is 
the assumption that in the earlier stage of heavy-ion collisions a 
thermally equilibrated fire cylinder is formed.  Although this is
a reasonable approximation as it gives a satisfactory explanation
for the observed particle yields and spectra from these collisions
\cite{braun}, a fully microscopic transport model including initial 
stage string dynamics is still desirable. Moreover, any `anomalous' 
phenomena in heavy-ion collisions have to be discussed with respect 
to proton-induced reactions. Both the CERES and the HELIOS-3 collaboration 
have also measured dilepton production in proton-nucleus reactions, it is
thus of interest to see if the data in proton-nucleus reactions can be
explained in a `conventional' way. This has not been possible in
Ref. \cite{li95b}, as the fire cylinder model is not applicable 
to proton-induced reactions.
 
In Refs. \cite{cass95a,cass95b}, the Hadron-String Dynamics \cite{cassing}
has been used to study dilepton production from both proton-nucleus
and nucleus-nucleus collisions. The conclusion from these studies
is that although conventional mechanisms are sufficient to account for
the dilepton spectra from proton-induced collisions, medium effects
as that proposed in Ref. \cite{li95b} are needed to explain the enhanced
low-mass dileptons in nucleus-nucleus collisions.  In this paper, we 
carry out a similar calculation as in Refs. \cite{li95b,cass95a,cass95b},
but use initial hadron phase space distributions generated by string 
fragmentation in the initial stage of the relativistic quantum molecular 
dynamics (RQMD) \cite{sorge89,sorge95a}. It is worthwhile to mention that
the RQMD model has been quite successful in describing many aspects of 
heavy-ion collisions at SPS/CERN energies.
 
This paper is organized as follows. In Section II, we discuss briefly
the string dynamics for the initial stage of heavy-ion collisions
and the hadronic rescattering in later stages. In Section III, we
present the formulae for dilepton production from hadronic interactions and
hadronic decay. The results for dilepton production in proton-nucleus
collisions are presented in Section IV. In section V, we compare our results
for S+Au and S+W collisions to the CERES and the HELIOS-3 data. The
predictions for dilepton production in Pb+Au collisions will be discussed 
in Section VI. Finally a brief summary and outlook is given in Section VII.

\section{the relativistic transport model}
 

The fragmentation of strings which may be viewed as an idealization
of longitudinally stretched chromoelectric flux-tubes is a
phenomenological approach 
to strong interactions
which has been applied
to  multi-particle production in the soft regime
for a long time \cite{lund1,lund2}.
String excitations are also an essential ingredient for
recent microscopic approaches to ultrarelativistic 
nucleus-nucleus collisions, like e.g.\ RQMD
\cite{sorge89,sorge95a}. 
Inelastic hadron-hadron collisions  are described by forming strings
from ingoing quarks and their primordial momenta.
This concept is generalized to nuclear collisions by allowing for
multiple collisons of a projectile with several target nucleons.
As a new feature present only with nuclei as collision partners,
RQMD includes the fusion of several overlapping strings into a
`rope', a flux-tube with larger than elementary color charge as
the source.
Created strings and ropes decay subsequently, because quark pairs
are created and screen the initial fields.
It is often assumed that
new hadrons are formed only after the two ingoing nuclei have
passed through each other.
This picture emerges also from the RQMD model at
sufficiently high beam energies, because the typical times for
hadron formation from string and rope fragmentation are getting
larger than the passage time of the two Lorentz contracted nuclei.


From the RQMD model (version 2.1) we obtain the  hadrons from 
the primary nucleon-nucleon collisions
in the initial stage of heavy-ion collisions. 
These hadrons and their 
distributions 
in position and momentum space
are then used as the input  to the relativistic transport 
model \cite{li95b}. When in-medium masses are used, both the yield 
and distribution of hadrons from string fragmentation may be different 
from that using the free masses \cite{cassing}. Since the hadron 
abundance and distribution in heavy-ion collisions reach the 
equilibrium values very quickly as shown in our calculations, the 
results obtained in the following are thus not sensitive to the 
change of this initial distributions.  We shall thus ignore such 
effects by using the same initial hadron abundance and spatial and 
momentum distributions whether the free or in-medium masses are used.  

The treatment of subsequent hadronic rescattering in the relativistic
transport model is similar to that in Ref. \cite{li95b}.  In the following, 
we mention briefly those processes that are directly relevant for 
dilepton production.  For a pair of pions with a total invariant mass 
$M$, a rho meson of this mass is formed with an isospin-averaged cross 
section given by the Breit-Wigner form \cite{li95a}
\begin{eqnarray}\label{ppr}
\sigma _{\pi \pi \rightarrow \rho } (M) = {8\pi\over {\bf k}^2}
{(m_\rho \Gamma _\rho )^2\over (M^2-m_\rho ^2)^2+(m_\rho \Gamma _\rho )^2}
\big({M\over m_\rho}\big)^2,
\end{eqnarray}
where ${\bf k}$ is the pion momentum in the center-of-mass frame of the
rho meson. Similarly, for $\pi\rho \rightarrow a_1$ the isospin-averaged 
cross sections \cite{xiong92} is
\begin{eqnarray}\label{prp}
\sigma _{\pi \rho \rightarrow a_1 } (M) = {4\pi\over 3{\bf k}^2}
{(m_{a_1} \Gamma _{a_1} )^2\over (M^2-m_{a_1} ^2)^2+(m_{a_1} \Gamma _{a_1})^2},
\end{eqnarray}
The kaon-antikaon collision mainly proceeds through the formation and 
decay of a phi meson, with an isospin-averaged cross section given by
\begin{eqnarray}\label{kkp}
\sigma _{K {\bar K} \rightarrow \phi } (M) = {3\pi\over {\bf k}^2}
{(m_\phi \Gamma _\phi )^2\over (M^2-m_\phi ^2)^2+(m_\phi \Gamma _\phi )^2}
\big({M\over m_\phi}\big)^2.
\end{eqnarray}
 
In treating meson decays we use the momentum-dependent decay width. For 
$\rho\rightarrow \pi\pi$ and $\phi\rightarrow K\bar K$, they are given, 
respectively, by \cite{li95a}
\begin{eqnarray}\label{wrho}
\Gamma _{\rho\rightarrow \pi\pi} (M)=
{g^2_{\rho\pi\pi}\over 4\pi} {\big(M^2-4m_\pi^2\big)^{3/2}\over 12 M^2},
\end{eqnarray}
and
\begin{eqnarray}\label{wphi}
\Gamma _{\phi\rightarrow K\bar K} (M)=
{g^2_{\phi K \bar K}\over 4\pi} {\big(M^2-4m_K^2\big)^{3/2}\over 6 M^2},
\end{eqnarray}
where $g_{\rho\pi\pi}^2/4\pi \approx 2.9$ and $g_{\phi K\bar K}^2/4\pi
\approx 1.7$ are determined from the measured width at $m_\rho$ and $m_\phi$,
respectively.
For the decay width of $a_1\rightarrow \pi\rho$, we use the result of 
Ref. \cite{xiong92}, i.e.,
\begin{eqnarray}\label{wa}
\Gamma _{a_1\rightarrow \pi\rho}={G_{a_1\pi\rho}^2 |{\bf k}|\over
24\pi m_{a_1}^2}\big[2(p_\pi\cdot p_\rho )^2
+m_\rho ^2(m_\pi^2+|{\bf k}|^2)\big],
\end{eqnarray}
where {\bf k} is the pion momentum in the rest frame of $a_1$;
$G_{a_1\pi\rho}\approx 14.8$ GeV$^{-1}$ is determined from the $a_1$ decay
width in free space using its centroid mass. There is no simple expression 
for the decay width of $\omega\rightarrow \pi^+\pi^-\pi^0$ \cite{krs84},
so we use the approximation that this width is proportional to the mass 
of omega meson. 
This approximation becomes exact in the chiral limit 
of $m_\pi\rightarrow 0$.
 
To study consistently the effects of dropping vector meson masses
on the dilepton spectrum in heavy-ion collisions, we need a model for
in-medium vector meson masses that can be incorporated into the
relativistic transport model. In Ref. \cite{li95b}, this is achieved by
extending the Walecka model \cite{qhd86} from the coupling of nucleons 
to scalar and vector fields to the coupling of light quarks to these fields,
using the ideas of the meson-quark coupling model \cite{thomas94}
and the constituent quark model. Here, we briefly review this model.
 

For a system of baryons (we take the nucleon as an example), pseudoscalar 
mesons ($\pi$ and $\eta$ mesons), vector mesons (rho and omega mesons), and
the axial-vector meson ($a_1$) at temperature $T$ and baryon density
$\rho _B$, the scalar field $\langle S \rangle$ is determined 
self-consistently from 
\begin{eqnarray}
m_S^2\langle S \rangle &=&{4g_S\over (2\pi )^3}\int d{\bf k} {m_N^*\over
E^*_N}\Big[{1\over \exp ((E^*_N-\mu _B)/T)+1}
+{1\over \exp ((E^*_N+\mu _B)/T)+1}\Big]\nonumber\\
&+&{0.45g_S\over (2\pi )^3}\int d{\bf k} {m_\eta^*\over E_\eta ^*}
{1\over \exp (E_\eta ^*/T)-1}+
{6g_S\over (2\pi )^3}\int d{\bf k} {m_\rho^*\over E_\rho ^*}
{1\over \exp (E_\rho ^*/T)-1}\nonumber\\
&+&{2g_S\over (2\pi )^3}\int d{\bf k} {m_\omega^*\over E_\omega ^*}
{1\over \exp (E_\omega ^*/T)-1}
+{6\sqrt 2 g_S\over (2\pi )^3}\int d{\bf k} {m_{a_1}^*\over E_{a_1}^*}
{1\over \exp (E_{a_1}^*/T)-1},
\end{eqnarray}
where we have used the constituent quark model relations for the nucleon
and vector meson masses \cite{thomas94}, i.e.,
\begin{eqnarray}
m_N^*=m_N-g_S\langle S\rangle ,~m_\rho^*\approx
m_\rho-(2/3)g_S\langle S\rangle ,
~m_\omega^*\approx m_\omega -(2/3)g_S\langle S\rangle ,
\end{eqnarray}
and the quark structure of the $\eta$ meson in free space which leads to
\begin{eqnarray}
m_\eta^*\approx m_\eta -0.45g_S\langle S\rangle ,
\end{eqnarray}
and the Weinberg sum rule relation between the rho-meson and $a_1$ meson
masses \cite{wein67,kapu94}, i.e,
\begin{eqnarray}
m_{a_1}^*\approx m_{a_1}-(2\sqrt 2/3)g_S\langle S\rangle .
\end{eqnarray}
In the calculation we use the scalar and vector coupling parameters 
of the original Walecka model that are fitted to the nuclear matter
properties at normal density. In high-energy heavy-ion collisions, 
the system is not necessary in themal and chemical equilibrium.
The thermal distributions in Eq. (7) are thus replaced by the 
hadron momentum distributions in determining the scalar field.

\section{dilepton production: formalism}
 
The main contributions to dileptons with mass below 1.2 GeV are the
Dalitz decay of $\pi^0$, $\eta$ and $\omega$, the direct leptonic decay of
vector mesons such as $\rho^0$, $\omega$ and $\phi$, and the pion-pion 
annihilation which proceeds through the $\rho^0$ meson, and the 
kaon-antikaon annihilation that proceeds through the $\phi$ meson.
 
The Dalitz decay of $\pi^0$, $\eta$, and $\omega$ contributes significantly
to dileptons with mass below 2$m_\pi$. The differential width of Dalitz 
decay is related to its radiative decay width \cite{land85}. For 
pseudoscalar meson $P$ (either $\pi^0$, $\eta^0$ or $\eta^\prime$), 
we have \cite{land85}
\begin{eqnarray}
{d\Gamma (P\rightarrow \gamma  l^+l^-)\over dM}&
=&{4\alpha\over 3\pi}{\Gamma (P\rightarrow 2\gamma)\over M}
\Big(1-{4m_l^2\over M^2}\Big)^{1/2}\nonumber\\
&\times& \Big(1+{2m_l^2\over M^2}\Big)
\Big(1-{M^2\over m_P^2}\Big)^3|F_P(M^2)|^2,
\end{eqnarray}
where $M$ is the mass of produced dilepton, $\alpha$ is the fine structure
constant, $\Gamma (P\rightarrow 2\gamma)$ is the measured radiative 
decay width of a pseudoscalar meson \cite{pada}, and $m_l$ is the mass 
of the lepton. In the case of dielectron production, $m_l=m_e\approx 
0.51$ MeV can be neglected. The electromagnetic form factors of $\pi^0$, 
$\eta$ and $\eta^\prime$ are parameterized, respectively, as \cite{land85}
\begin{eqnarray}
F_{\pi^0}(M^2)=1+b_{\pi^0}M^2,
\end{eqnarray}
\begin{eqnarray}\label{etaff}
F_{\eta}(M^2)=\Big(1-{M^2\over \Lambda_{\eta}^2}\Big)^{-1},
\end{eqnarray}
\begin{eqnarray}\label{etapff}
F_{\eta^\prime}(M^2)={m_\rho ^4\over (M^2-m_\rho ^2)^2
+(m_\rho \Gamma_\rho)^2},
\end{eqnarray}
where $b_{\pi^0}=5.5$ GeV$^{-2}$, $\Lambda_{\eta}\approx 0.72$ GeV,
and $m_\rho$ and $\Gamma_\rho$ are the mass and width of rho meson,
respectively. These form factors describe reasonably well the empirical 
data \cite{land85}.
 
For $\omega\rightarrow \pi^0 l^+l^-$, the differential decay width is
given by
\begin{eqnarray}
{d\Gamma (\omega\rightarrow \pi^0 l^+l^-)\over dM}&
=&{2\alpha\over 3\pi}{\Gamma (\omega\rightarrow \pi^0\gamma)\over M}
\Big(1-{4m_l^2\over M^2}\Big)^{1/2}\Big(1+{2m_l^2\over M^2}\Big)
\nonumber\\
&\times&\Big[\big(1+{M^2\over m_\omega^2-m_{\pi^0}^2}\big)^2
-\big({2m_\omega M\over m_\omega^2-m_\pi^2}\big)^2\Big]
|F_{\omega}(M^2)|^2,
\end{eqnarray}
where $\Gamma (\omega\rightarrow \pi^0 \gamma )=0.717$ MeV is the omega
meson radiative decay width \cite{pada}. In Ref. \cite{land85}, the 
electromagnetic form factor was parameterized as
\begin{eqnarray}\label{omeffl}
F_{\omega}(M^2)=\Big(1-{M^2\over \Lambda_{\omega}^2}\Big)^{-1},
\end{eqnarray}
with $\Lambda_{\omega}=0.65$ GeV, somewhat smaller than $m_\rho$, indicating
the deviation from the vector dominance model. 
 
The treatment of $a_1$ Dalitz decay is slightly different. Since in
our dynamical model the processes $a_1\leftrightarrow\pi\rho$ and 
$\rho\rightarrow l^+l^-$ are treated explicitly, we have already 
included that part of $a_1$ contribution to dileptons which proceeds 
through a physical $\rho$ meson as a two-step process.  Thus, in 
evaluating the $a_1$ Dalitz decay ($a_1\rightarrow \pi l^+l^-$)
we do not need to introduce the vector-dominance model form factor.
Otherwise, there would be double counting.
 
The direct leptonic decay of vector mesons is another important source of 
dileptons. For dileptons of mass in the region of interest to this work,
we consider mainly the leptonic decay of $\rho ^0$,  $\omega$ and $\phi$ 
mesons.  The decay width for $\rho ^0\rightarrow l^+l^-$ is given by
\begin{eqnarray}\label{ree}
\Gamma _{\rho^0\rightarrow  l^+l^-}(M)&=&
{g_{\rho\gamma}^2e^2\over M^4}{M\over 3}
\Big(1-{4m_l^2\over M^2}\Big)^{1/2}\Big(1+{2m_l^2\over M^2}\Big)\nonumber\\
&=&{\alpha^2\over \big(g_{\rho\pi\pi}^2/4\pi \big)} {m^4_\rho \over 3M^3}
\Big(1-{4m_l^2\over M^2}\Big)^{1/2}\Big(1+{2m_l^2\over M^2}\Big),
\end{eqnarray}
where $M^4$ in the denominator arises from the virtual photon propagator
and $M$ in the numerator comes from the phase space integration. In obtaining
the second expression, we have used the vector dominance relation 
$g_{\rho\gamma}=em_\rho^2/g_{\rho\pi\pi}$ \cite{bh88}. Using 
$g_{\rho\pi\pi}^2/4\pi\approx 2.9$ in Eq.(\ref{ree}), we get 
$\Gamma _{\rho^0\rightarrow e^+e^-}\approx 5$ keV, and this is somewhat 
different from the measured width of 6.5 keV, indicating the breaking 
of universal vector coupling. In the calculation, we use instead
\begin{eqnarray}
\Gamma _{V\rightarrow e^+e^-}(M) 
=C_{e^+e^-}{m_V^4\over M^3}, 
\end{eqnarray}
\begin{eqnarray}
\Gamma _{V\rightarrow\mu^+\mu^-}(M) 
=C_{\mu^+\mu^-}{m_V^4\over M^3} 
\Big(1-{4m_\mu^2\over M^2}\Big)^{1/2}
\Big(1+{2m_\mu^2\over M^2}\Big),
\end{eqnarray}
where $V$ stands for $\rho^0$, $\omega$, or $\phi$. The coefficient 
$C_{e^+e^-}$ in the dielectron channel is $8.814\times 10^{-6}$,
$0.767\times 10^{-6}$, and $1.344\times 10^{-6}$ for $\rho^0$, $\omega$,
and $\phi$ decay, respectively, which are determined from the measured
width \cite{pada}. Similarly, from the measured width of decaying into 
dimuons, we determine the coefficient $C_{\mu^+\mu^-}$ to be 
$9.091\times 10^{-6}$ and $1.071\times 10^{-6}$ for $\rho^0$ 
and $\phi$ decay, respectively.  For $\Gamma _{\omega\rightarrow 
\mu^+\mu^-}$, only an upper bound is given in the particle data book 
\cite{pada}; we assume that this width is the same as that for the 
dielectron channel. Our fit to HELIOS-3 data for p+W collisions shows 
that this is a good approximation (see Fig. 3 below).
 
The formation of an omega meson from the three-pion interaction is neglected.
In this calculation, all omega mesons are thus from the initial
stage of heavy-ion collisions through string fragmentation.
For rho and phi mesons we include also secondary processes such as
pion-pion annihilation and kaon-antikaon annihilation, in addition
to primary production from string fragmentation. The contribution
of pion-pion and kaon-antikaon annihilation to dilepton production,
which have been known to be important for heavy-ion collisions, is
therefore treated as a two-step process with the explicit intermediate 
vector meson formation, propagation, and decay. It can be easily shown that
the product of the rho meson formation cross section, Eq. (\ref{ppr}),
and the branching ratio, $\Gamma _{\rho^0\rightarrow  l^+l^-}/\Gamma _\rho$,
for the rho meson to decay into dilepton leads to a dilepton production 
cross section that is the same as that from pion-pion annihilation 
in the usual form factor approach
\cite{gale87,li95a}, i.e.,
\begin{eqnarray}\label{form}
\sigma_{\pi^+\pi^-\rightarrow \rho^0\rightarrow  l^+l^-}=
{8\pi\alpha^2 k\over 3M^3} {m_\rho^4\over (M^2-m_\rho^2)^2+(m_\rho
\Gamma_\rho)^2}  
\Big(1-{4m_l^2\over M^2}\Big)^{1/2}
\Big(1+{2m_l^2\over M^2}\Big).
\end{eqnarray}
The same relation holds for dileptons from kaon-antikaon annihilation.
The interaction of a pion and a rho meson dominantly produces an $a_1$, 
and their contribution to dilepton production is thus included in 
our model through $a_1$ Dalitz decay.
 
It should be mentioned that when medium effects on vector meson masses are
included, $m_{\rho}$, $m_\omega$, and $m_\phi$ in the above expressions
are replaced by $m_\rho^*$, $m_\omega^*$, and $m_\phi^*$, respectively.
Also, the in-medium decay widths are used in these expressions, and
they are calculated from Eqs. (\ref{wrho}-\ref{wa}) with in-medium masses.
As in Ref. \cite{li95b}, we have neglected the collisional broadening of 
vector meson widths in medium \cite{hag95}, based on the argument that 
their magnitudes are comparable to the mass resolution in CERES 
experiments, so they do not affect appreciably the final results.

In our model, dileptons are emitted continuously during the time 
evolution of the colliding system. The way the dilepton yield is 
calculated can be illustrated by rho meson decay. Denoting, at time t, 
the differential multiplicity of neutral rho mesons by $dN_{\rho ^0} 
(t)/dM$, then the differential dilepton production probability is given by
\begin{eqnarray}\label{sum}
{dN_{l^+l^-}\over dM} =\int _0^{t_f} {dN_{\rho^0} (t)\over dM} 
\Gamma _{\rho^0
\rightarrow l^+l^-}(M) dt + {dN_{\rho ^0} (t_f)\over dM} 
{\Gamma _{\rho^0\rightarrow l^+l^-} (M)\over\Gamma _\rho (M)},
\end{eqnarray}
where $t_f$ is the freeze-out time, which is found to be about 20 fm/c.
The first term corresponds to dilepton emission before freeze out while
the second term is from decay of rho mesons still present after freeze out.
 
\section{dilepton production: proton-nucleus reactions}
 
As already mentioned in the introduction, any `anomalous' phenomena in
heavy-ion collisions need to be compared with proton-induced reactions. 
In order to claim that the CERES and HELIOS-3 dilepton data as indications 
of medium effects in heavy-ion collisions, it is necessary that we can 
describe the dilepton data in proton-nucleus collisions with the 
`conventional' mechanism. In these reactions, dileptons are expected 
to be produced outside the nucleus due to finite formation time, so 
no `exotic' phenomena are expected to happen when compared to dilepton 
production from the proton-proton interaction.
 
The results for dilepton spectra from p+Be collisions at 450 GeV are shown
in Fig. 1, together with the data from the CERES collaboration \cite{ceres95}.
It is seen that the data can be well reproduced by the Dalitz decay of 
$\pi^0$, $\eta$ and $\omega$ mesons, and the direct leptonic decay of
$\rho^0$, $\omega$ and $\phi$ mesons. These results are thus similar 
to that found in Refs. \cite{cass95a} using the Hadron-String Dynamics, 
and are also similar to the `cocktail' constructed by the CERES 
collaboration from known and expected sources of dileptons \cite{ceres95}.
We see that very low mass dileptons are mainly from the Dalitz decay 
of $\pi^0$ and are strongly suppressed by experimental acceptance cuts 
in the opening angle and transverse momentum. Dileptons in the mass 
region from 0.15 to 0.45 GeV are mainly from $\eta$ meson Dalitz decay. 
Around $m_{\rho ,\omega}$, the contribution from $\omega$ decay is 
more important than that from $\rho^0$ decay, as the latter has a very 
broad mass distribution.  Dileptons with mass around 1 GeV are mainly 
due to $\phi$ meson decay.
 
A similar comparison with the CERES data for p+Au collisions at 450 GeV
is given in Fig. 2. Again, the data are very well reproduced by
conventional sources of Dalitz decay and direct vector meson decay.
Comparing Fig. 1 with Fig. 2 we see that both the experimental data and
our theoretical results are essentially the same in the two collisions,
so there is no discernible difference in the dilepton spectra in 
going from an extremely light Be target (which is similar to proton-proton 
interactions) to a much heavier Au target.
 
The comparison with the HELIOS-3 dimuon data from p+W collisions at 200 GeV
is presented in Fig. 3.  Up to dimuon mass of about 1.3 GeV, the data are 
again very well explained by the Dalitz decay of $\eta$ and $\omega$ and 
the direct vector meson decay of $\rho^0$, $\omega$ and $\phi$. The 
low mass dimuons are mainly from the $\eta$ Dalitz decay, while the 
$\omega$ meson Dalitz decay is important in the mass range of 0.4 to 0.6 GeV.
As in the CERES case, the contribution from the $\omega$ decay is more 
important than that from the $\rho ^0$ around $m_{\rho ,\omega}$.
Both the CERES and the HELIOS-3 data for dilepton production in 
proton-nucleus reactions are thus consistent with each other, in 
spite of quite different experimental setups and acceptance cuts.
 
\section{dilepton production: nucleus-nucleus collisions}
 
Comparing the experimental data from nucleus-nucleus collisions to those from
proton-nucleus collisions, both the CERES and the HELIOS-3 collaboration
have found substantial enhancement of low-mass dileptons in the mass region
from about $2m_\pi$ to about 0.6 GeV, which cannot be explained by
uncertainties and errors of the normalization procedure \cite{tser95}.
In going from proton-induced reactions to heavy-ion collisions, the 
main difference is additional contributions from pion-pion and 
kaon-antikaon annihilation to dilepton production in heavy-ion collisions. 
If the observed enhancement of low-mass dileptons can be explained by 
pion-pion annihilation, it is fine but cannot be considered as an 
`anomaly', as the importance of pion-pion annihilation has already be 
seen in the DLS data \cite{dls88}. However, as shown in Refs. 
\cite{li95b,cass95a,cass95b} and will be shown in this Section, including
the contribution from pion-pion annihilation but without any medium 
effects, the theoretical results in the low-mass region are still 
significant below both the CERES and HELIOS-3 data.

\subsection{CERES: S+Au collisions}
 
We first show in Fig. 4 the initial pion and proton rapidity distributions 
and transverse momentum spectra obtained in the RQMD for central S+Au 
collisions at 200 GeV/nucleon and impact parameter $b\le 3$ fm, 
corresponding to a final charged particle 
multiplicity of $dN_{ch}/d\eta\approx 125$ in the rapidity range 
$2.1<\eta<2.65$ as in the CERES experiments. Also shown is the time 
evolution of central baryon density. For comparison, the initial conditions
and time evolution of the central density used in the fire cylinder 
model of Ref. \cite{li95b} are also shown. It is seen that the two 
initial conditions agree reasonably with each other. Initially, the 
slope parameter of the pion transverse mass distribution is about 
165 MeV, the average baryon density in the central region is about
2.3 $\rho_0$, and the average energy density is about 2.5 GeV/fm$^3$.
In Fig. 5, we show the time evolution of the abundance of both participant 
baryons and produced mesons. We note that initially the pion number 
is about twice the rho meson number, and the omega meson number is 
about 1/3 of the rho meson number.  In the free meson mass case, rho 
mesons are seen to continuously decay, leading to an increase of the 
pion number. The number of higher baryon resonances, whose in-medium masses 
are always used in the relativistic transport model, increases initially 
as their number is below the equilibrium one. After reaching the 
equilibrium value, it starts to decrease as the system expands.  When 
in-medium meson masses are introduced in the model, the time evolution 
of the baryon abundance remains similar as their in-medium masses are 
essentially the same. However, the time evolution of the rho meson 
number changes to the one like the higher baryon resonances, indicating 
that initially their number is below the equilibrium value due to a 
smaller mass in medium.

After following the time evolution of the colliding system using the
relativistic transport model, hadronic observables can be calculated.
In Fig. 6, we show the final rapidity distributions of negatively-charged 
hadrons (mainly $\pi^-$ and $K^-$) and protons together with the 
experimental data from the NA35 collaboration \cite{na35} and the 
NA44 collaboration \cite{na44}.  Compared with that in Ref. \cite{li95b}, 
we see a significant improvement of agreements between the theoretical 
results with the data in the backward rapidity region. The comparison 
of transverse mass spectra of protons and pion in mid- to forward-rapidity 
region with experimental data for a similar system S+Pb from the NA44 
collaboration \cite{na44} is shown in Fig. 7. Overall, the theoretical 
results agree well with the data.
 
Also of interest is the $\eta / \pi^0$ ratio in these collisions. The 
WA80 collaboration has recently published their measurement of this 
quantity for minimum-biased events \cite{wa95}. Some very preliminary 
data with large error bars were reported in Ref. \cite{wa93} for 
central collisions. The comparison of our theoretical results with 
the WA80 data is shown in Fig. 8. Our $\eta /\pi^0$ ratio is larger 
than the minimum-biased data (solid circles), but somewhat smaller 
than the central collision data (solid squares), especially in the 
low transverse momentum region. We believe that our results correspond 
reasonably to the ratio in CERES experiments, as less central events 
are selected in CERES experiments than in WA80 experiments, which 
has an average charged-particle multiplicity density of more than 160 
in the same rapidity region \cite{wa93,wa92}.
 
With free meson masses, the calculated dilepton spectra, normalized by the
average charged-particle multiplicity, are shown in Fig. 9 together with 
the CERES data. As in the case of proton-nucleus reactions, dileptons 
with mass below 0.1 GeV are mainly from $\pi^0$ Dalitz decay. 
The $\eta$ Dalitz decay is important in the mass range from 0.15 to 
0.35 GeV.  From 0.35 to 0.7 GeV, dileptons are mainly from pion-pion 
annihilation through $\rho^0$ meson.  Contrary to the observation
in proton-nucleus reactions where the $\omega$ decay is more
important around $m_{\rho ,\omega}$, in S+Au collisions the contribution
from $\rho^0$ decay becomes more important when free meson masses are used.
This is mainly due to an enhanced rho contribution from pion-pion annihilation
in heavy-ion collisions. A bump around 0.7 GeV in the dilepton spectrum
from $\eta^\prime$ meson decay is a result of the vector meson dominance
form factor in Eq. (\ref{etapff}). There should have appeared a similar bump 
in the contribution from $a_1$ Dalitz decay if we had included the 
vector meson dominance form factor rather than treated it as a 
two-step process in the rho meson mass region.
 
The importance of pion-pion annihilation in the low-mass region, as
emphasized by the CERES collaboration \cite{ceres95}, can be more
clearly seen in Fig. 10, where the CERES data are compared
with theoretical results calculated with and without the contribution
from pion-pion annihilation.  Without pion-pion annihilation, the 
theoretical results for dileptons with invariant mass from 0.3 to 0.65 GeV
are significantly reduced, and the disagreement with the experimental 
data becomes comparable to that found by the CERES collaboration based 
on known and expected sources \cite{ceres95}.
 
The inclusion of pion-pion annihilation, though found to be quite 
important in the mass region of interest, still does not give enough 
number of dileptons in the mass region from 0.25 to 0.6 GeV. Furthermore, 
for dileptons with mass around $m_{\rho,\omega}$ there are more dileptons
predicted by the theoretical calculations than shown in the experimental 
data. This is very similar to our earlier results based on a thermally 
equilibrated fire cylinder model \cite{li95b}. These results are
also similar to those of Cassing {\it et al} \cite{cass95a} based on
the Hadron-String Dynamics model and Srivastava {\it et al} \cite{gale95}
based on the hydrodynamical model, as shown in Fig. 11. We note that a
strong peak around $m_\phi$ in the results of Srivastava {\it et al} 
will become a bump once the mass resolution of the CERES collaboration is
properly included.
 
The failure of these models with free meson masses in explaining
the CERES data has led to the suggestion that medium modifications
of vector meson masses are needed as shown in Refs.
\cite{li95a,cass95a,cass95b}.
The comparison of our results obtained with in-medium meson masses with
the CERES data is shown in Fig. 12.  We have about a factor of 2-3 
enhancement of the dilepton yield in the mass range from 0.2 to 0.6 GeV, 
as compared with the results obtained with free meson masses. Overall, 
the agreement with the CERES data becomes much better with the use 
of in-medium vector meson masses. These results are again very similar 
to those in Ref. \cite{li95b} obtained in the fire-cylinder model.
 
\subsection{HELIOS-3: S+W collisions}

The same model has been used to calculate the dimuon spectra from 
central S+W collisions by the HELIOS-3 collaboration. The results obtained
with free meson masses are shown in Fig. 13.  As in proton-induced 
reactions, dimuons with mass below 0.3 GeV are from $\eta$ meson Dalitz 
decay. From 0.35 to 0.55 GeV, the $\omega$ meson Dalitz decay is 
important. In heavy-ion collisions, the contribution from $\rho^0$ 
decay becomes important due to pion-pion annihilation. However, the 
role of pion-pion annihilation in HELIOS-3 experiments is less significant 
than that in CERES experiments, as HELIOS-3 measures dileptons in the 
forward rapidity region which has a smaller charged-particle multiplicity 
than that in the mid-rapidity region measured by CERES.

With free meson masses, the theoretical results are below the HELIOS-3 
data in the mass region from 0.35 to 0.6 GeV, and slightly above the data
around $m_{\rho ,\omega}$. Qualitatively, this is similar to the situation
in the CERES case, but quantitatively, the discrepancy between the theory 
and the data is somewhat smaller in the HELIOS-3 case. For example, for 
the CERES data the theoretical results underpredicts most significantly 
around 0.4 GeV and are below the data by about a factor of 4, while
for the HELIOS-3 data this happens around 0.5 GeV, and the theoretical
prediction is below the data by less than a factor of two. In other 
words, the enhancement of low-mass dileptons is less pronounced in 
the HELIOS-3 than in the CERES experiments. This is again due to the 
fact that the HELIOS-3 measures dileptons in the forward rapidity region 
with a smaller charged-particle multiplicity than in the CERES experiments.
 
Our results obtained with in-medium meson masses are shown in Fig. 14.
Again, we see enhanced dilepton yield in the low mass region, and
a reduction around $m_{\rho ,\omega}$, as compared to the results
obtained with free meson masses. This brings the theoretical results
in better agreement with the data.  The importance of in-medium meson 
masses in explaining the HELIOS-3 data has also been found by 
Cassing {\it et al} \cite{cass95b}.
 
For dimuon spectra in p+W collision up to 1.3 GeV, the HELIOS-3
data can be well described by the Dalitz decays and direct vector meson
decay. For S+W collisions, the HELIOS-3 data above 1.2 GeV are grossly
underestimated by theoretical calculations with both free and in-medium 
vector meson masses. The enhancement of dilepton production in 
heavy-ion collisions in this intermediate mass region indicates that
thermal processes such as $\pi a_1 \rightarrow l^+l^-$ \cite{song94} 
and the decay of heavier vector mesons such as
$\omega (1390)$ \cite{gale94,haglin95} might become important.
Also, contributions from the quark-gluon plasma and the initial 
Drell-Yan processes may not be negligible.
 
\section{dilepton production: predictions for Pb+Au collisions}
 
Dilepton production in Pb+Au collisions at 160 GeV/nucleon is currently
being measured by the CERES collaboration. Based on the model outlined
above, which has been successful in explaining both the CERES and HELIOS-3
data with reduced vector meson masses in hot and dense matter, we 
present our predictions for dilepton spectra from this collision using 
current CERES mass resolution and acceptance cuts for S+Au collisions.
 
We first consider central Pb+Au collisions.  In Fig. 15, we show the 
initial pion and proton rapidity and transverse mass distributions as 
well as the time evolution of baryon density in this collision. Initially, 
the pion transverse slope parameter is about 190 MeV, the average 
baryon density is about 4 $\rho_0$, and the average energy density is 
about 3.3 GeV/fm$^3$.  These initial parameters are somewhat higher 
than in S+Au collisions at 200 GeV/nucleon. As in Ref. \cite{li95b}, 
the system expands slightly slower in the case of dropping meson masses. 
The time evolution of the abundance of participant baryons and produced 
mesons is shown in Fig. 16 for both free and in-medium meson masses.  
As in S+Au collisions, the initial pion number is about twice of that 
of rho mesons, and the omega number is about 1/3 of the rho meson number.  
For the case of free meson masses, the time dependence of the particles 
is similar to what we see in the S+Au collision. Introducing in-medium 
meson masses gives a similar time dependence of the meson abundance 
as in the S+Au collision with in-medium masses. However, a peculiar 
behavior of the baryon abundance occurs at about 3 fm/c after the expansion,
when baryon resonances are reformed after initial decays, and this is 
not seen in S+Au collisions.  We believe that this is due to the 
larger $\pi/N^*$ ratio in Pb+Au collisions ($\approx 15$ at t=3 fm/c) 
than in S+Au collisions ($\approx 4$ at the same time), which thus 
favors the formation of baryon resonances in the Pb collision.
 
For central Pb+Au and Pb+Pb collisions some preliminary data for 
the charged-particle rapidity distribution and transverse mass spectra
have become available from the NA49 and the NA44 collaboration.
We compare in Fig. 17 the calculated negatively-charged particle 
(mainly $\pi^-$ and $K^-$) rapidity distribution with preliminary data 
from the NA49 collaboration \cite{na35}. The agreement with the data 
is fairly good. Also shown is Fig. 17 is the final proton rapidity 
distribution, which shows more appreciable stopping than in S+Au 
collisions. The nucleon to pion ratio in the mid-rapidity is still 
about 0.15 as in S+Au collisions. The final proton rapidity distribution 
is somewhat different from the initial one from string fragmentation 
(see upper left panel of Fig. 15). Because of longitudinal expansion, 
a hole is seen to develop in the mid-rapidity. We  note that in our 
initial proton rapidity distribution hyperons have not been included, 
while our final proton rapidity distribution (lower panel of Fig. 17)
includes also the protons from hyperon decay.  The calculated final 
proton and pion transverse momentum spectra agree with the preliminary 
data from the NA44 collaboration \cite{na44} as shown in Fig. 18.
 
The theoretical predictions for the dielectron spectra in central 
Pb+Au collisions with CERES mass resolution and acceptance cuts are 
given in Fig. 19(a) for the two scenarios of free meson masses and 
in-medium meson masses. The normalization factor $dN_{ch}/d\eta$ here 
is the average charge particle pseudo-rapidity density in the 
pseudo-rapidity range 2 to 3, and is about 440 in this collision.
With free meson masses, we see a strong peak around $m_{\rho ,\omega}$, 
which is dominated by $\rho^0$ meson decay as a result of an enhanced
contribution from pion-pion annihilation in Pb+Au collisions than
in S+Au and proton-nucleus collisions. With in-medium meson masses, 
the $\rho$ meson peak shifts to a lower mass, and the peak around 
$m_{\rho ,\omega}$ becomes a shoulder arising mainly from $\omega$ 
meson decay. At the same time we see an enhancement of low-mass dileptons 
in the region of 0.25-0.6 GeV as in S+Au collisions.
 
Since the maximum baryon density reached in heavy-ion collisions
decreases with the impact parameter, we expect a decrease of medium
effects on the dilepton spectrum at large impact parameters.
Experimentally, this effect can be studied by measuring the dependence
of the dilepton spectrum on the charged particle multiplicity. We have 
thus calculated the dilepton spectrum from Pb+Au collisions at 160 
GeV/nucleon for a number of impact parameters.  In Fig. 20, we show the 
impact parameter dependence of the initial average baryon density 
$\rho /\rho _0$, the final total negatively-charged particle multiplicity 
$N_{ch}^-$, and the average charge particle pseudorapidity 
density at midrapidity $dN_{ch}^-/d\eta$. Apparently, both 
decrease with increasing impact parameter; the initial baryon density 
goes from about 4$\rho _0$ at b=0 fm to 1.9$\rho _0$ at 9 fm, and 
$N_{ch}^{-}$ decreases from about 820 at b=0 fm to about 206 at b=9 fm.


For both Dalitz decay and vector meson decay, the ratio of total 
number of dileptons (without acceptance cut)
to the total negatively-charged hadron multiplicity 
$N_{ch}^-$ as a function of $N_{ch}^-$ are shown in Fig. 21.
It is seen that in both free and in-medium meson mass cases, the
dilepton yield from Dalitz decay increases linearly with the charge
particle multiplicity, so the normalized yield is almost a constant.
On the other hand, the dilepton yield from vector meson decay increases 
more than linearly due to contributions from pion-pion and kaon-antikaon 
annihilation. If we express $N_{ee}\propto (N_{ch}^-)^\alpha$, then our 
results show that $\alpha\approx 1.3$ for the case of free meson masses
and $\alpha\approx 1.5$ for the case of in-medium meson masses. The 
stronger dependence on the charged particle multiplicity in the latter 
case is due to the stronger medium effects in central collisions.

Finally, the calculated dilepton spectra from Pb+Au collisions at
three different impact parameters of b= 3, 6, and 9 fm are shown in 
Fig. 19(b), 19(c),and 19(d).  The enhancement of low-mass dileptons 
are seen to decrease with increasing impact parameter.

\section{summary and outlook}
 
In summary, we have studied in detail dilepton production from both
proton-nucleus and nucleus-nucleus collisions using the relativistic 
transport model with initial conditions determined by the string 
fragmentation from the initial stage of the RQMD model.
 
We have found that the dilepton spectra in proton-nucleus reactions
measured by the CERES and the HELIOS-3 collaboration can be well 
understood in terms of conventional mechanisms of Dalitz decay and
direct vector meson decay.
 
For dilepton spectra in central S+Au and S+W collisions, these
conventional mechanisms, however, fail to explain the data,
especially in the low-mass region from about 0.25 to about 0.6 GeV in
CERES experiments, and from 0.35 to 0.65 GeV in HELIOS-3 experiments.
Including the contribution from pion-pion annihilation, which is known 
to be important in the mass region from $2m_\pi$ to $m_{\rho ,\omega}$, 
removes some of the discrepancy.  But the data in the low mass region 
are still substantially underestimated, and that around $m_{\rho ,\omega}$ 
somewhat overestimated by theoretical calculations.  The agreement with 
the data is significantly improved when reduced in-medium vector meson 
masses are taken into account.  The results of the present study based 
on initial conditions from the RQMD model is thus very similar to our 
earlier results assuming that initially there is a thermally equilibrated 
fire-cylinder.
 
We have also presented predictions for the dilepton spectra from central 
Pb+Au collisions. With an increased pion density in Pb+Au than in 
Sulfur-induced collisions, the dilepton yield from Dalitz decay and 
$\omega$ decay increases roughly linearly with the charged-particle
multiplicity, whereas the contribution from pion-pion annihilation
increases more than linearly. This leads to some differences between 
the dilepton spectra from Pb+Au and Sulfur-induced collisions. We have 
also studied the impact parameter dependence of dilepton production in 
Pb+Au collisions, and found that the enhancement of low-mass dilepton 
decreases with increasing impact parameter. It would be very interesting 
to carry out experiments in which different centrality bins are 
selected so that this impact parameter dependence can be tested.

\vskip 1cm
 
We thank W. Cassing, A. Drees, I. Kralik, M. Murray, I. Tserruya, 
and T. Ullrich, and P. Wurm for helpful communications.
We are also very grateful to K. Wolf for providing us computing
resources and for useful discussions. GQL and CMK were supported 
by the National Science Foundation under Grant No. PHY-9509266, 
and GEB and HS were supported by the Department of Energy under 
Grant No. DE-FG02-88ER40388.

\pagebreak

\begin{figure}
\begin{center}
\epsfig{file=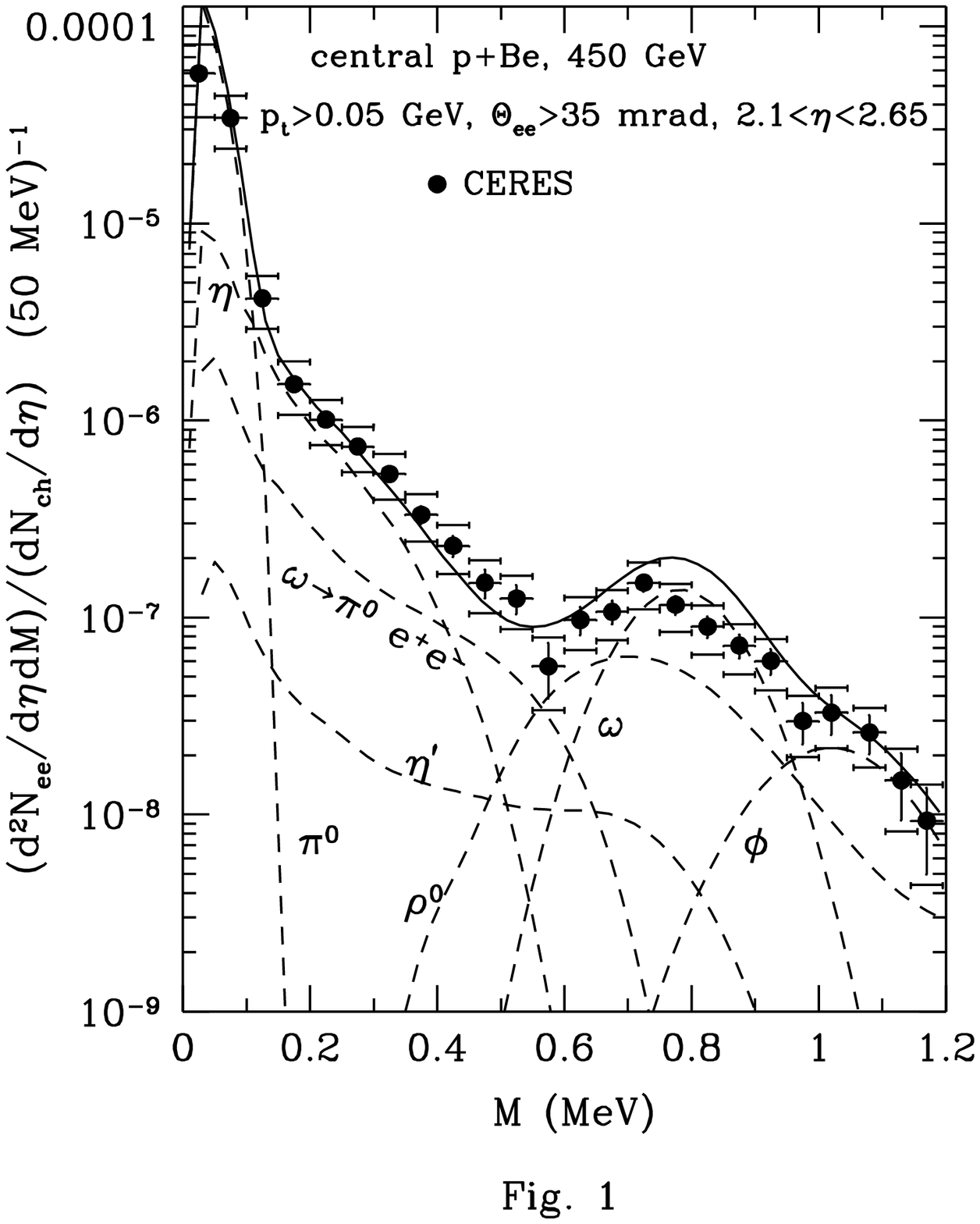,height=6in,width=5in}
\end{center}
\caption{Dilepton invariant mass spectra from p+Be collisions 
at 450 GeV after including the experimental acceptance cuts and mass 
resolution. Dashed curves give the dilepton spectra from different sources.
Experimental data from the CERES collaboration [34] are
shown by solid circles, with the statistical errors given by
bars. Brackets represent the square root of the quadratic
sum of systematic and statistical errors.}
\end{figure}

\newpage
 
\begin{figure}
\begin{center}
\epsfig{file=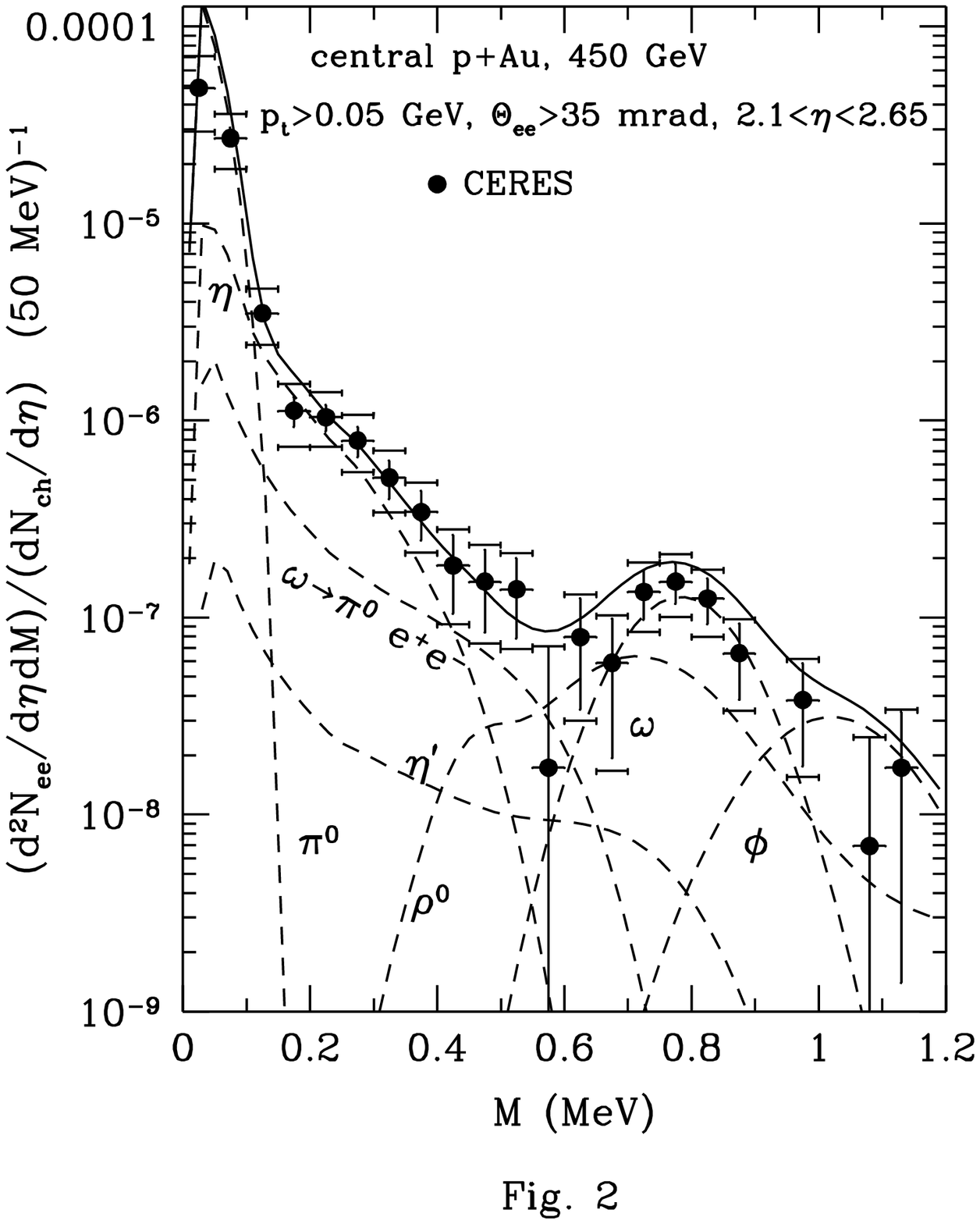,height=6in,width=5in}
\end{center}
\caption{Same as Fig. 1 for p+Au collisions.}
\end{figure}

\newpage
 
\begin{figure}
\begin{center}
\epsfig{file=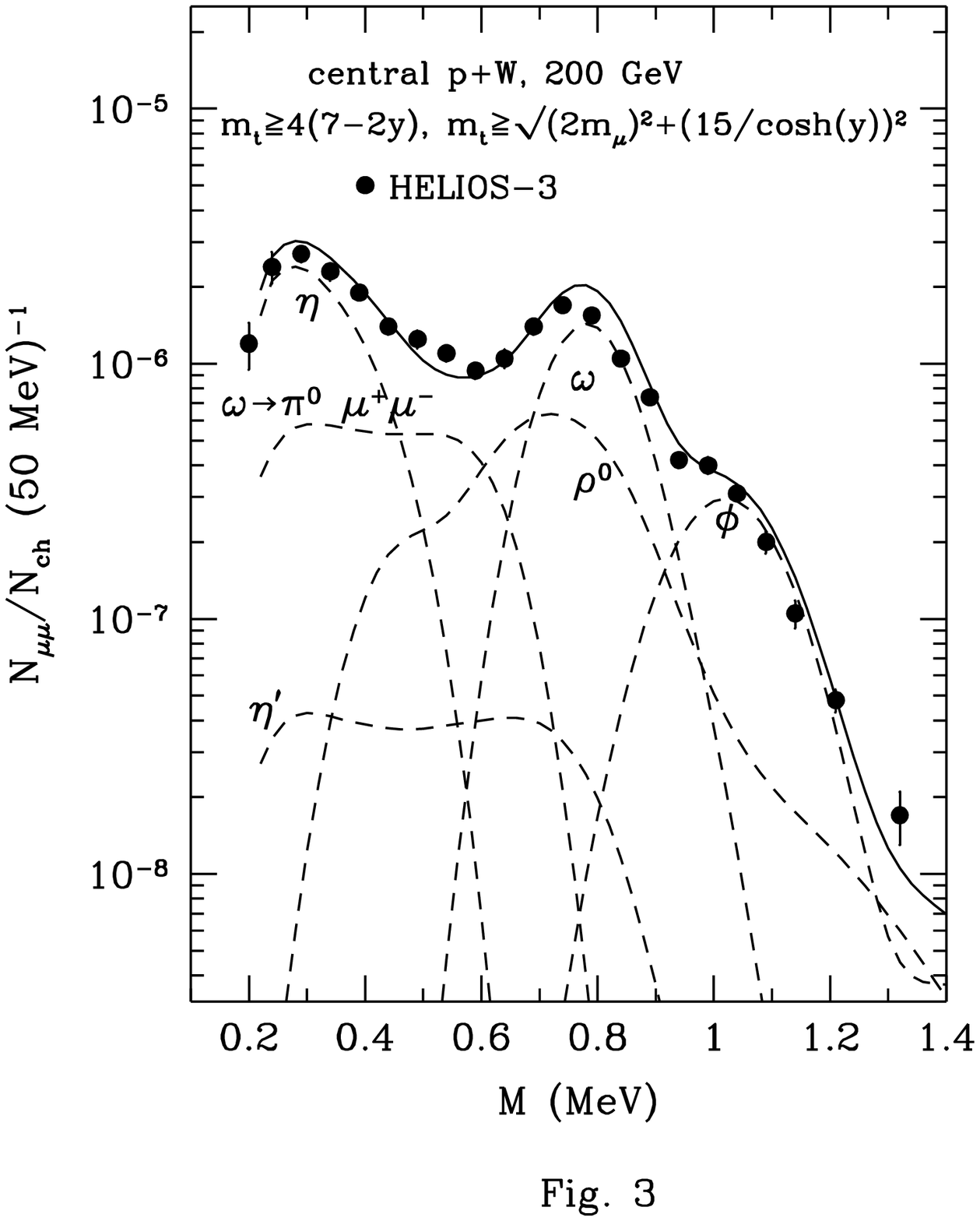,height=6in,width=5in}
\end{center}
\caption{Dilepton invariant mass spectra from p+W collisions 
at 200 GeV after including the experimental acceptance cuts and mass 
resolution. Dashed curves give the dilepton spectra from different sources.
Experimental data from the HELIOS-3 collaboration [35] are
shown by solid circles.}
\end{figure}

\newpage
 
\begin{figure}
\begin{center}
\epsfig{file=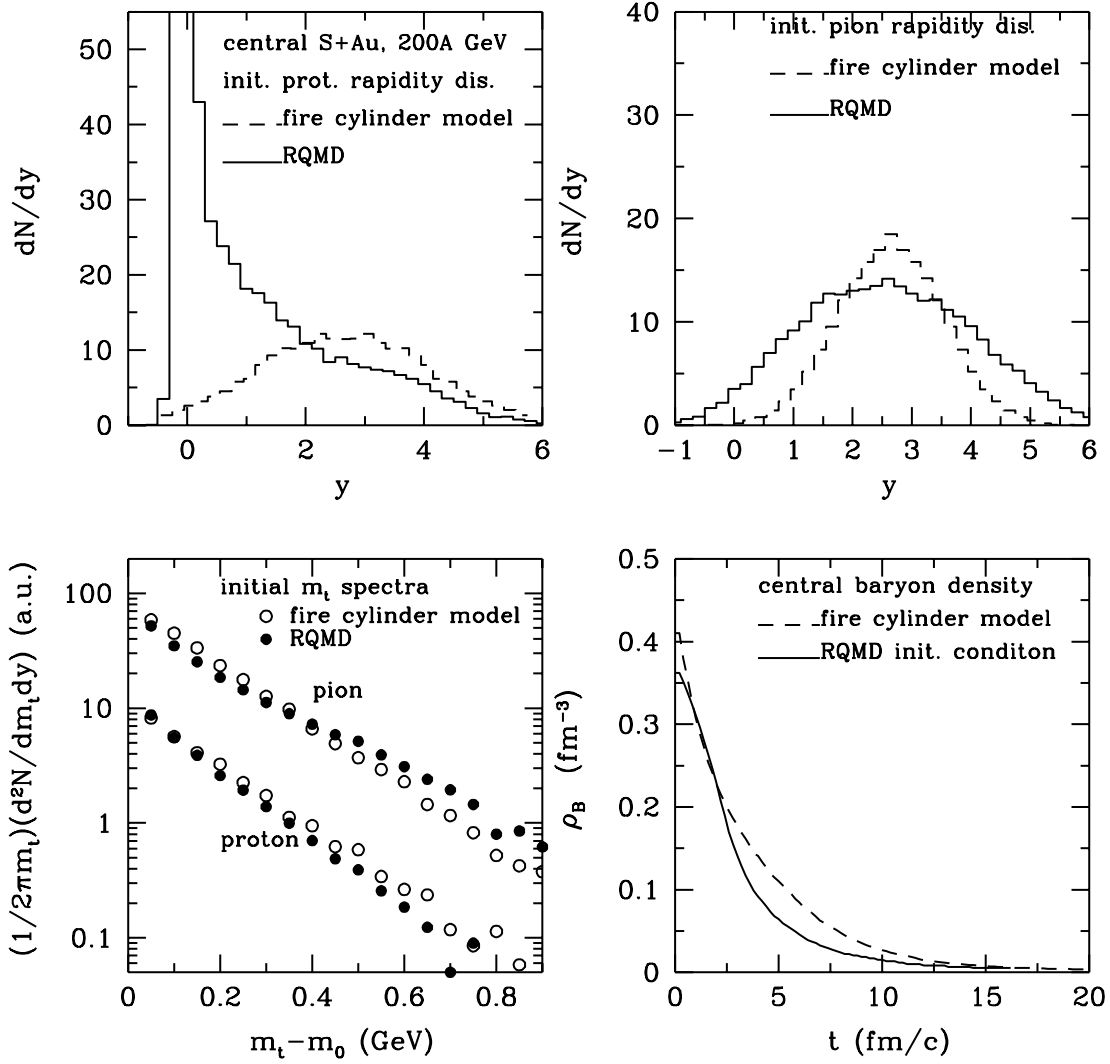,height=6in,width=6in}
\end{center}
\caption{The initial proton and pion rapidity and transverse mass
distributions from both the RQMD and the fire-cylinder model of
[36] for S+Au collisions at 200 GeV/nucleon.  Also shown is 
the time evolution of the baryon density using the two initial conditions.}
\end{figure}
 
\newpage

\begin{figure}
\begin{center}
\epsfig{file=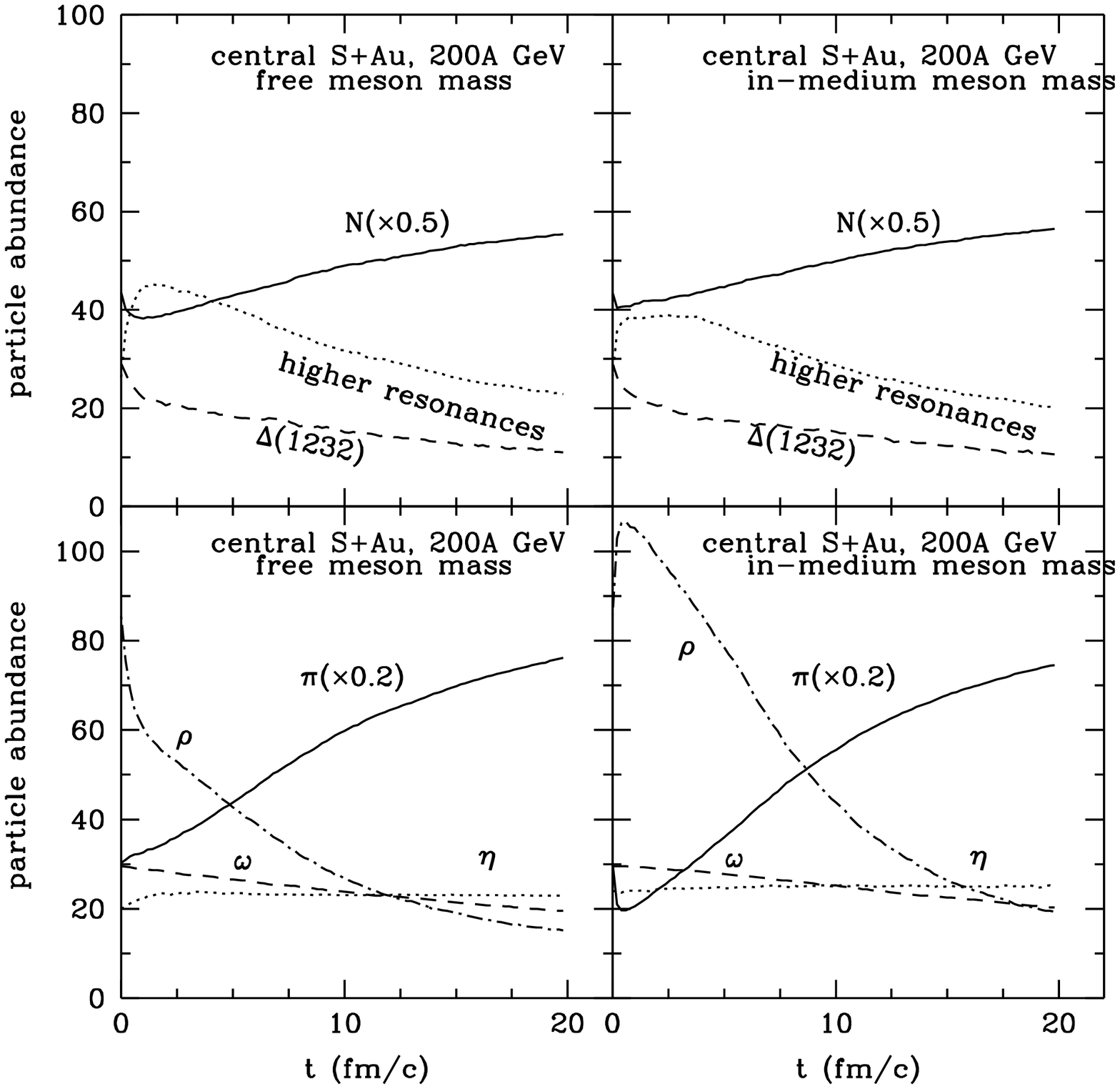,height=6in,width=6in}
\end{center}
\caption{The time evolution of hadron abundance in central
S+Au collisions at 200 GeV/nucleon.}
\end{figure}
 
\newpage

\begin{figure}
\begin{center}
\epsfig{file=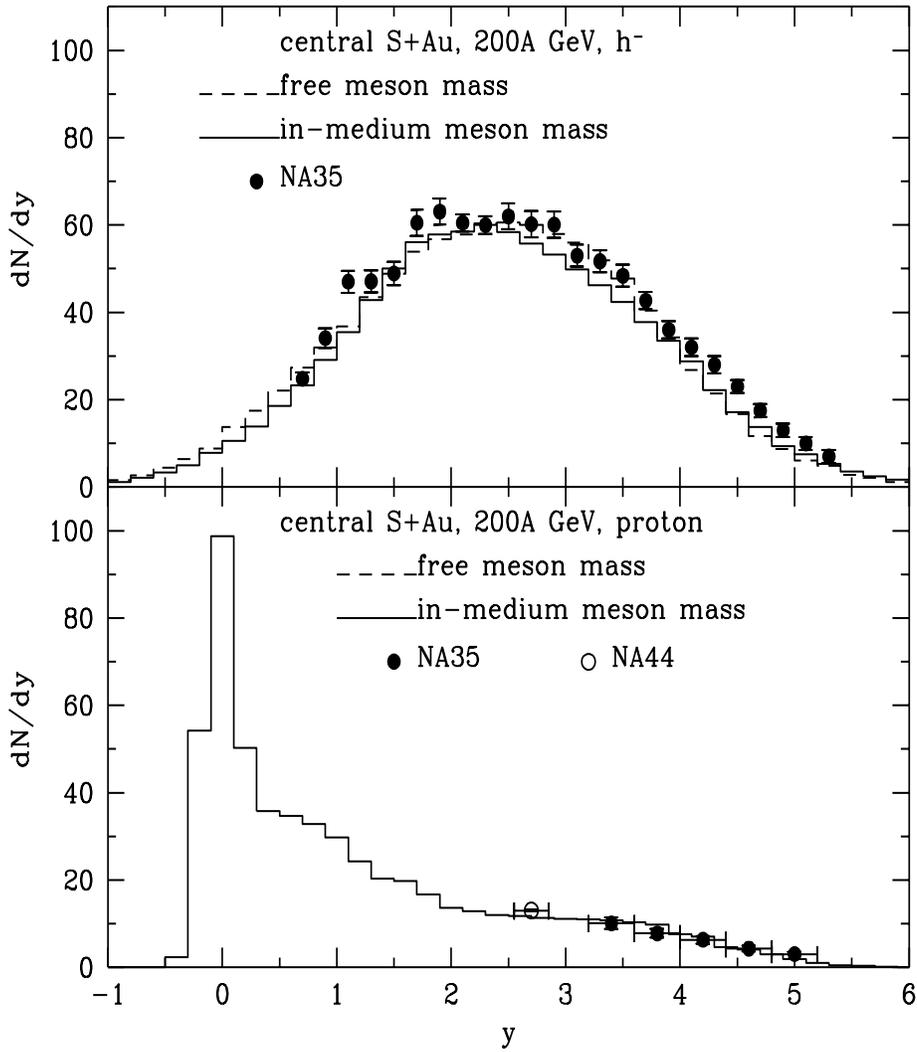,height=6in,width=5in}
\end{center}
\caption{The rapidity distributions of negatively-charged hadrons
and protons. Experimental data from the NA35 collaboration [61] 
are shown by solid circles, and that from NA44 [62] are shown 
by open circle.}
\end{figure}

\newpage
 
\begin{figure}
\begin{center}
\epsfig{file=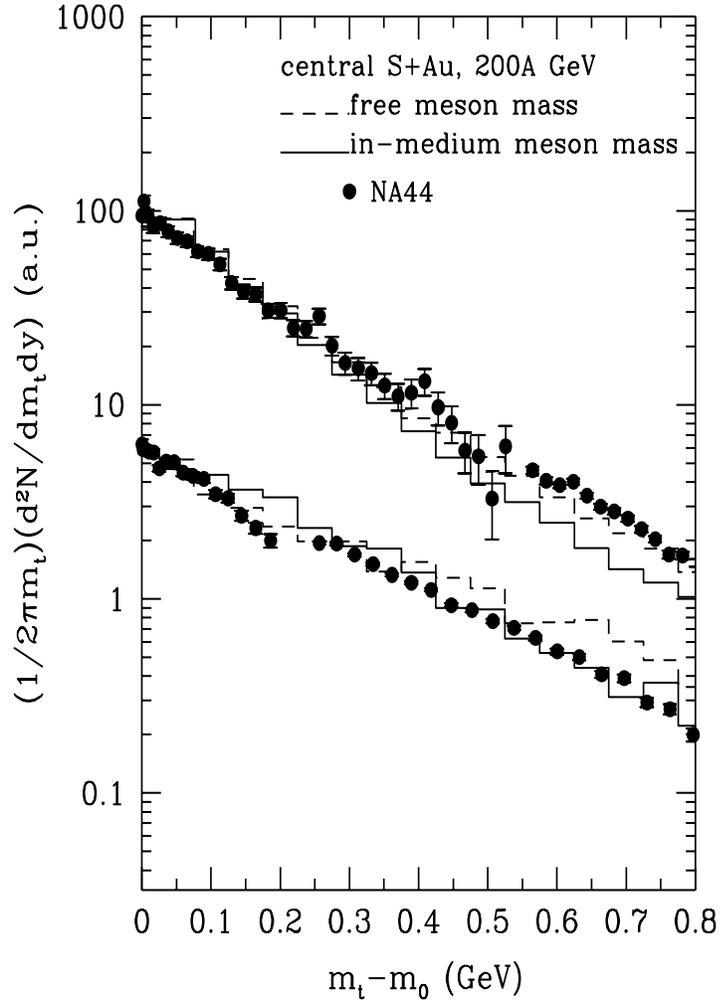,height=6in,width=5in}
\end{center}
\caption{The transverse mass spectra of protons and pions in the mid-
to forward-rapidity region. Experimental data from central S+Pb collisions 
by the NA44 collaboration [62] are shown by solid circles.}
\end{figure}
 
\newpage

\begin{figure}
\begin{center}
\epsfig{file=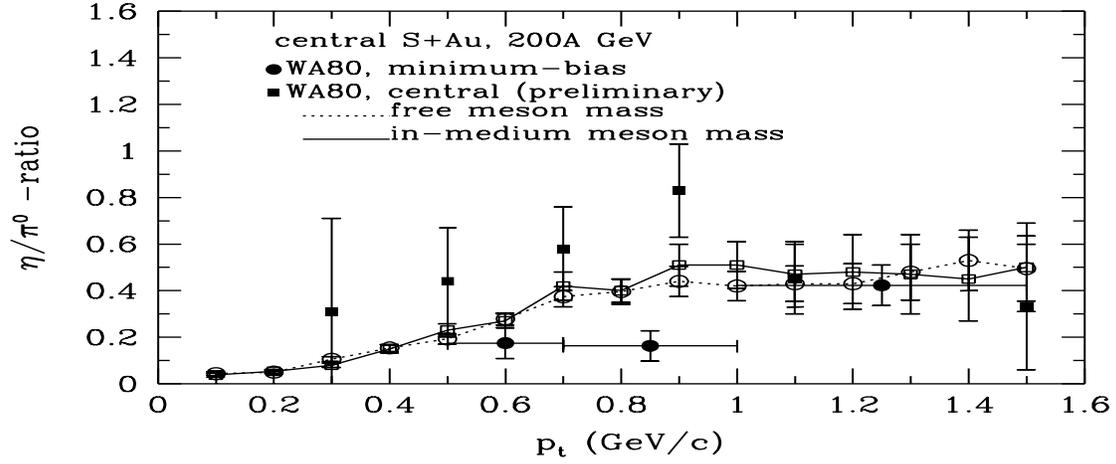,height=4in,width=6in}
\end{center}
\caption{The $\eta /\pi^0$ ratio as a function of transverse momentum.
Open circles and squares are theoretical results with free and
in-medium meson masses, respectively. Solid circles and squares
are the WA80 data for minimum-biased and central collisions, respectively.}
\end{figure} 

\newpage

\begin{figure}
\begin{center}
\epsfig{file=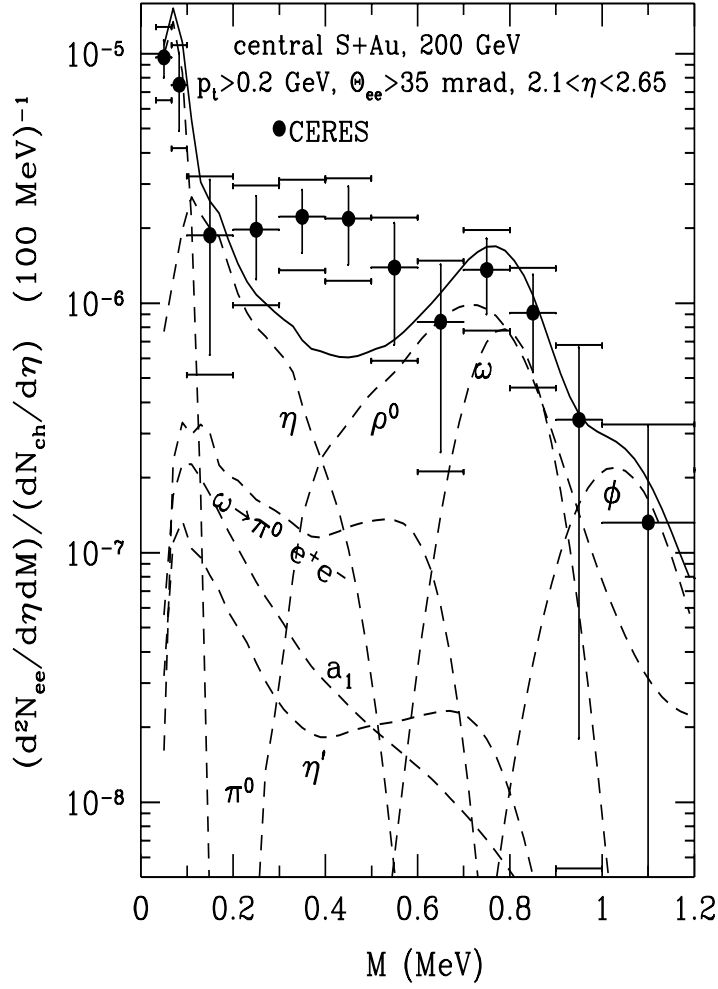,height=6in,width=5in}
\end{center}
\caption{Dilepton invariant mass spectra from S+Au collisions 
at 200 GeV using free meson masses and after including the experimental 
acceptance cuts and mass resolution. Dashed curves give the dilepton 
spectra from different sources. Experimental data from the CERES 
collaboration [34] are shown by solid circles, with the 
statistical errors given by bars. Brackets represent the square root 
of the quadratic sum of systematic and statistical errors.}
\end{figure}

\newpage

\begin{figure}
\begin{center}
\epsfig{file=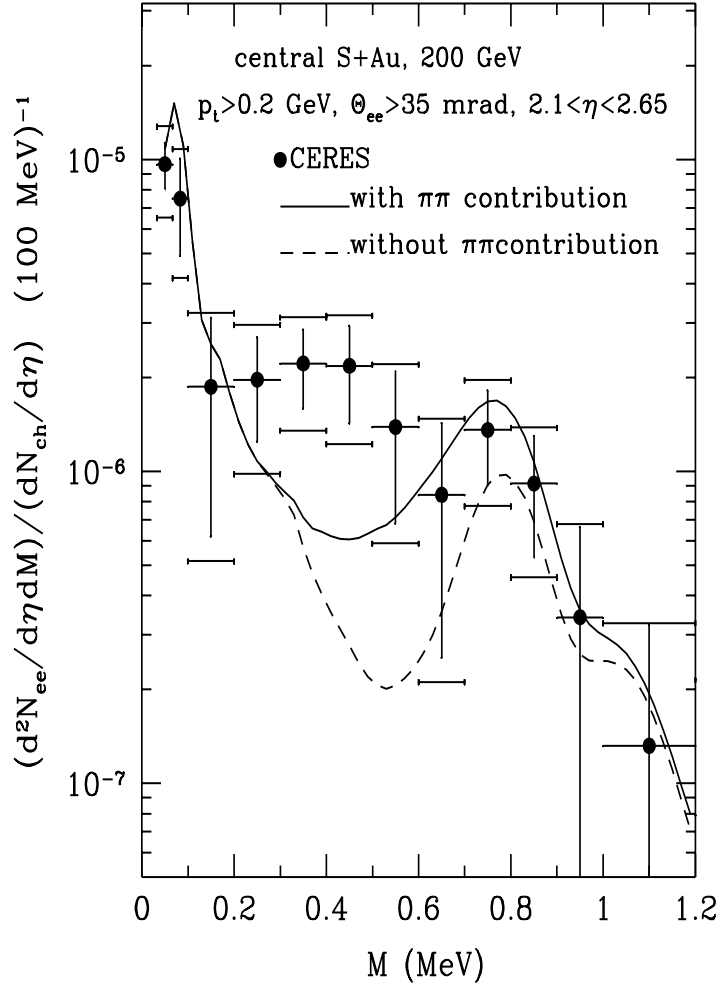,height=6in,width=5in}
\end{center}
\caption{The importance of pion-pion annihilation contribution
to dilepton spectra.}
\end{figure}

\newpage
 
\begin{figure}
\begin{center}
\epsfig{file=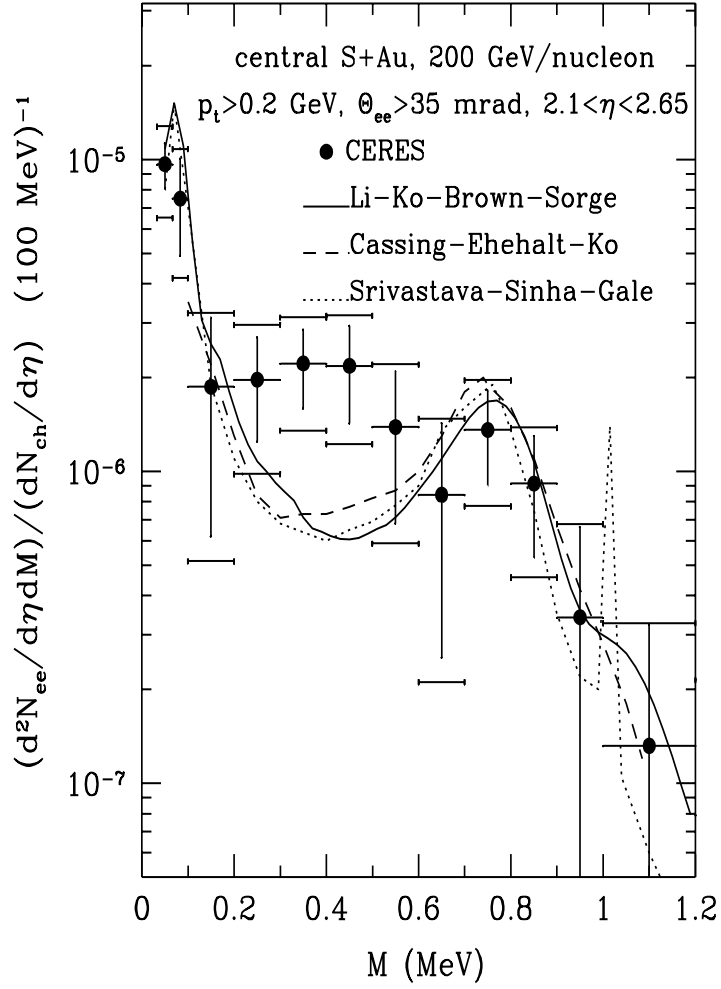,height=6in,width=5in}
\end{center}
\caption{Comparisons of dilepton spectra in central S+Au collisions
from three different model calculations. The solid curve is from this work,
the dashed curve is from Ref. [37], and the dotted curve is 
from Ref. [39].}
\end{figure}
 
\newpage

\begin{figure}
\begin{center}
\epsfig{file=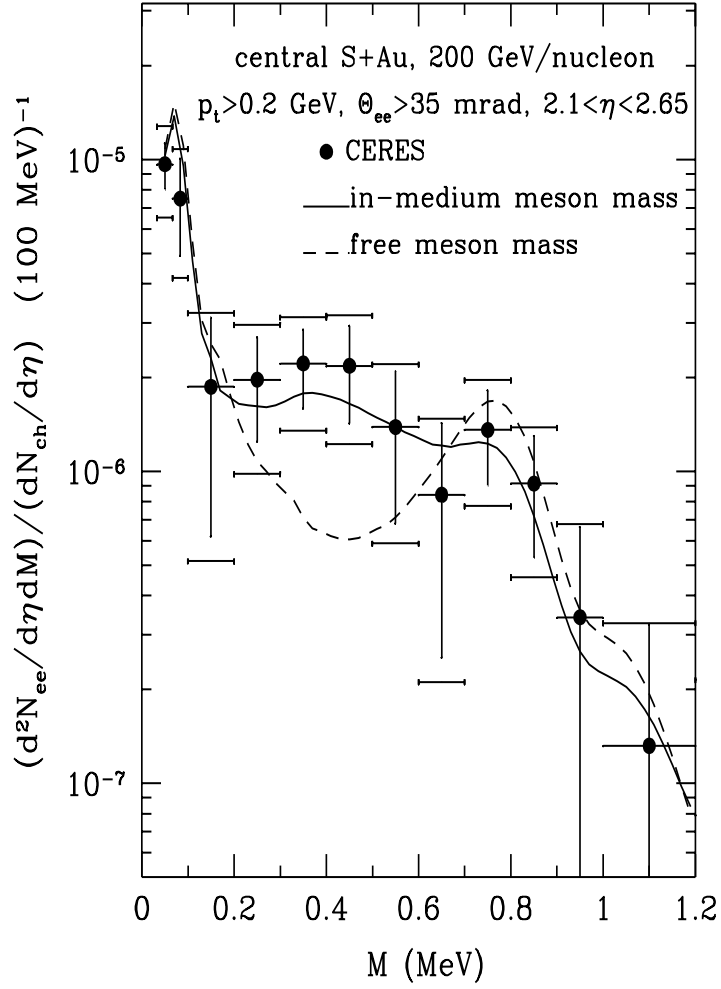,height=6in,width=5in}
\end{center}
\caption{Dilepton invariant mass spectra from central S+Au collisions 
with free (dashed curve) and in-medium (solid curve) meson masses.}
\end{figure}
 
\newpage

\begin{figure}
\begin{center}
\epsfig{file=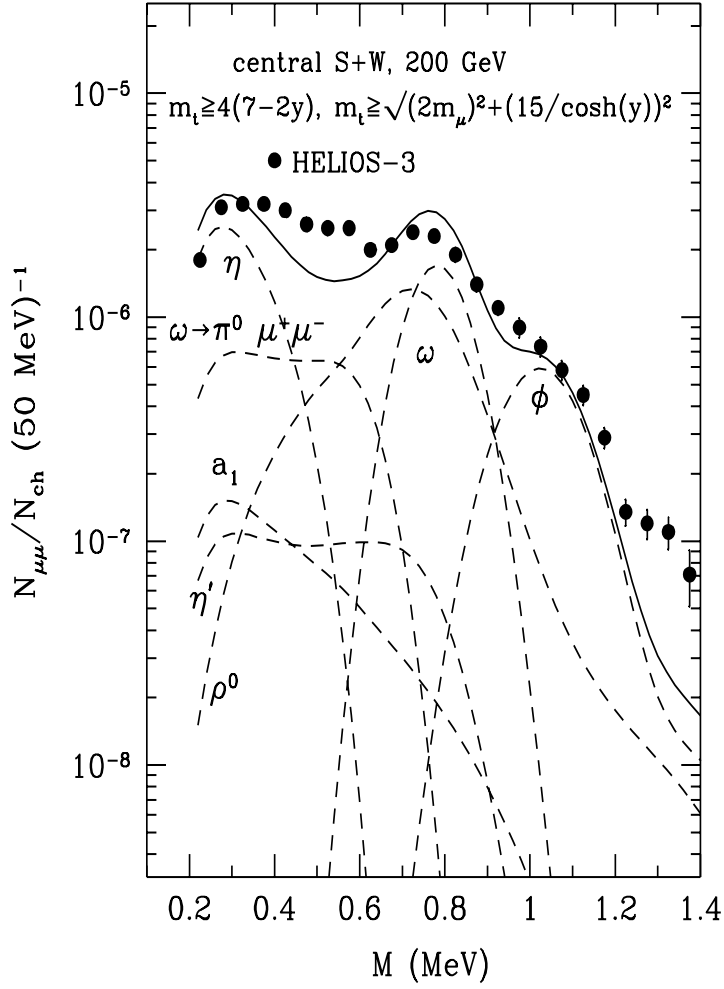,height=6in,width=5in}
\end{center}
\caption{Dilepton invariant mass spectra from S+W collisions 
at 200 GeV using free meson masses and after the including experimental 
acceptance cuts and mass resolution. Dashed curves give the contributions 
from different sources. Experimental data from the HELIOS-3 collaboration 
[35] are shown by solid circles.}
\end{figure}
 
\newpage

\begin{figure}
\begin{center}
\epsfig{file=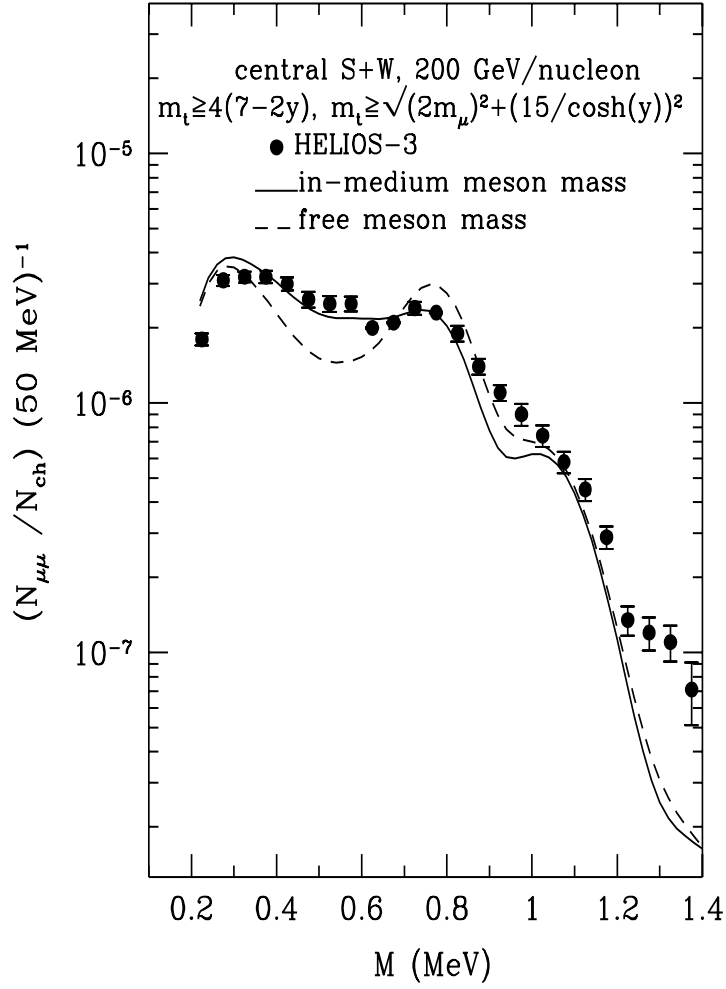,height=6in,width=5in}
\end{center}
\caption{Dilepton invariant mass spectra from S+W collisions 
at 200 GeV with free (dashed curve) and in-medium (solid curve) 
meson masses.}
\end{figure} 
 
\newpage

\begin{figure}
\begin{center}
\epsfig{file=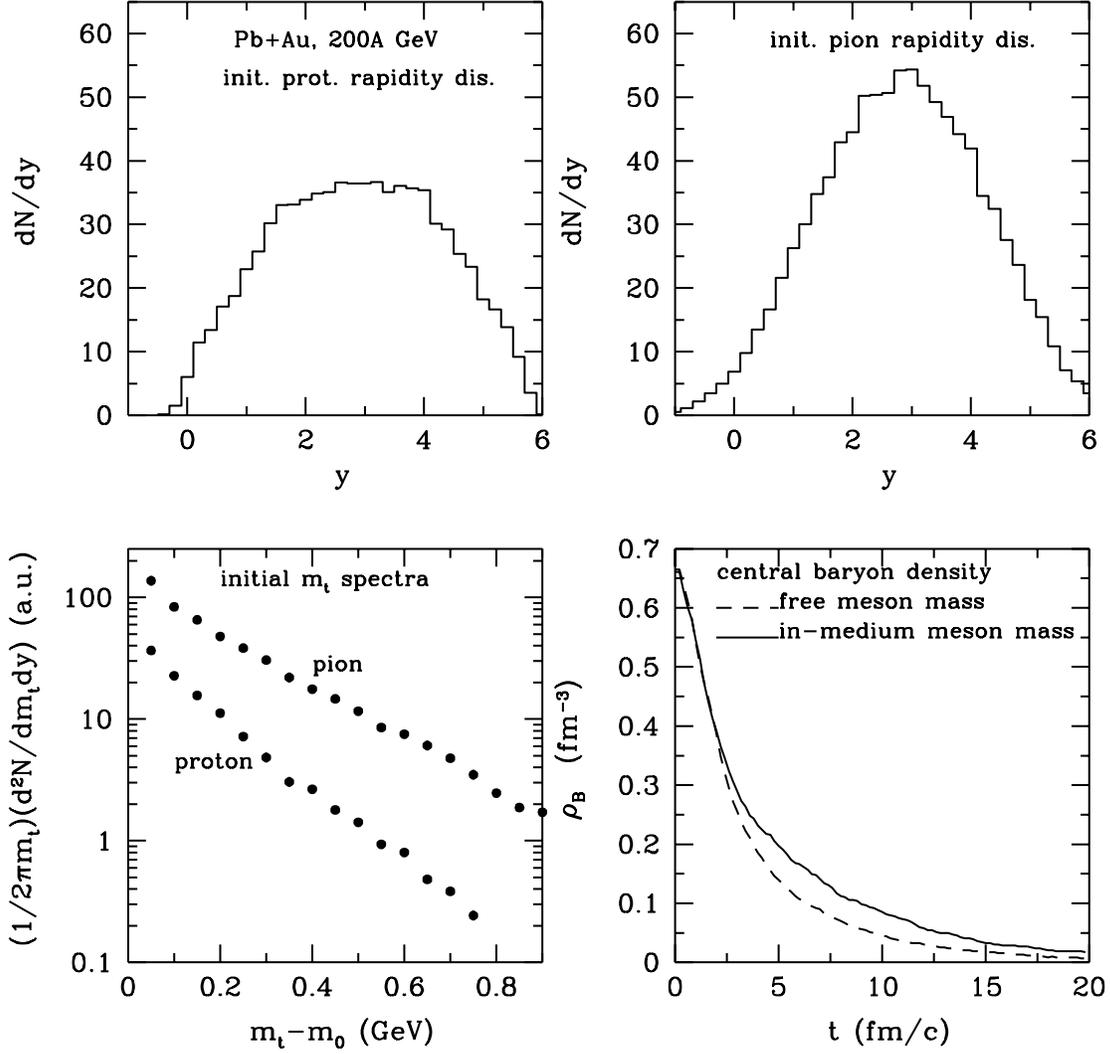,height=6in,width=6in}
\end{center}
\caption{The initial proton and pion rapidity and transverse mass
distributions in central Pb+Au collisions at 160 GeV/nucleon. Also shown 
is the time evolution of the baryon density.}
\end{figure}
 
\newpage

\begin{figure}
\begin{center}
\epsfig{file=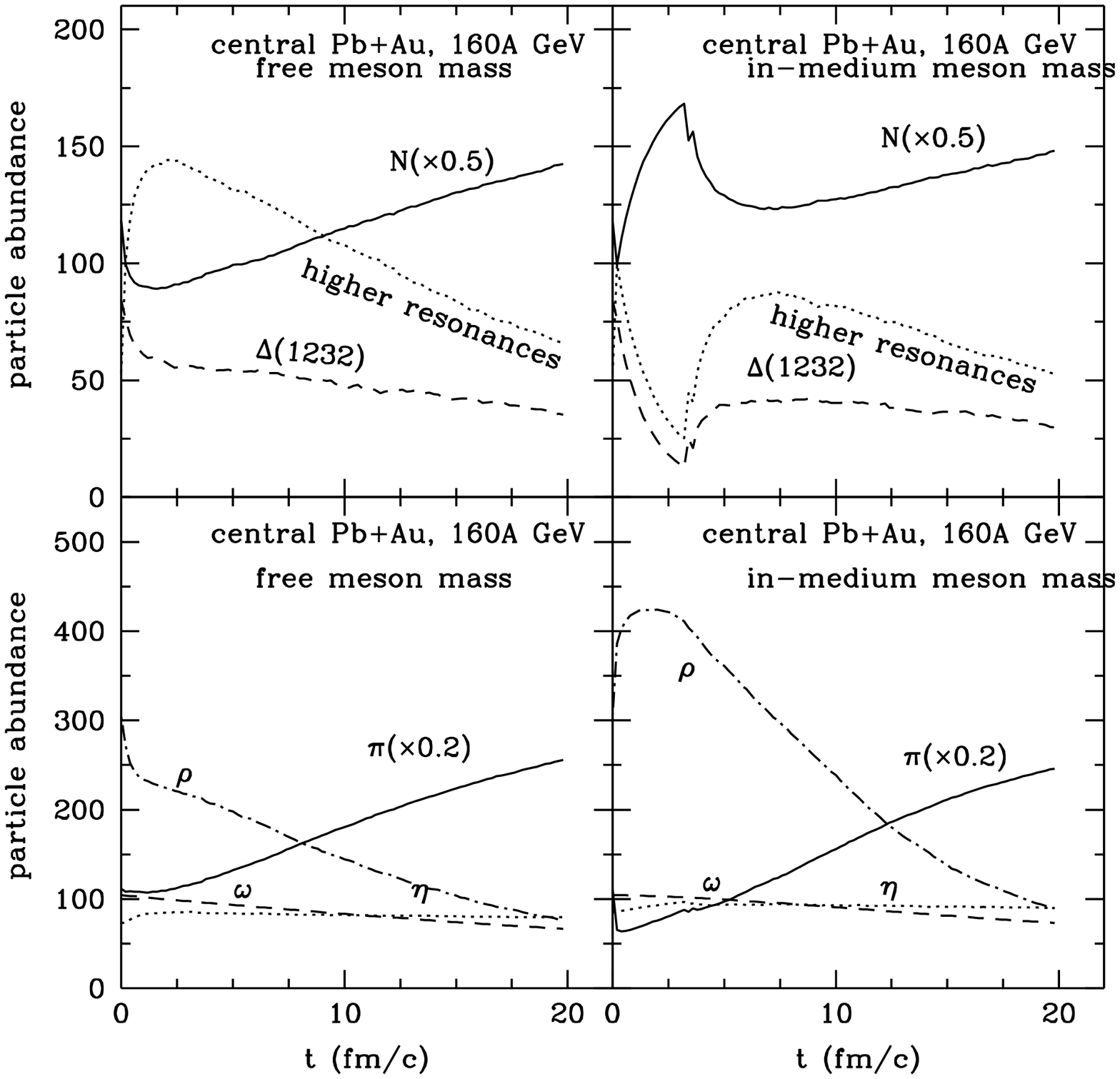,height=6in,width=6in}
\end{center}
\caption{The time evolution of hadron abundance in central
Pb+Au collisions at 160 GeV/nucleon.}
\end{figure}
 
\newpage

\begin{figure}
\begin{center}
\epsfig{file=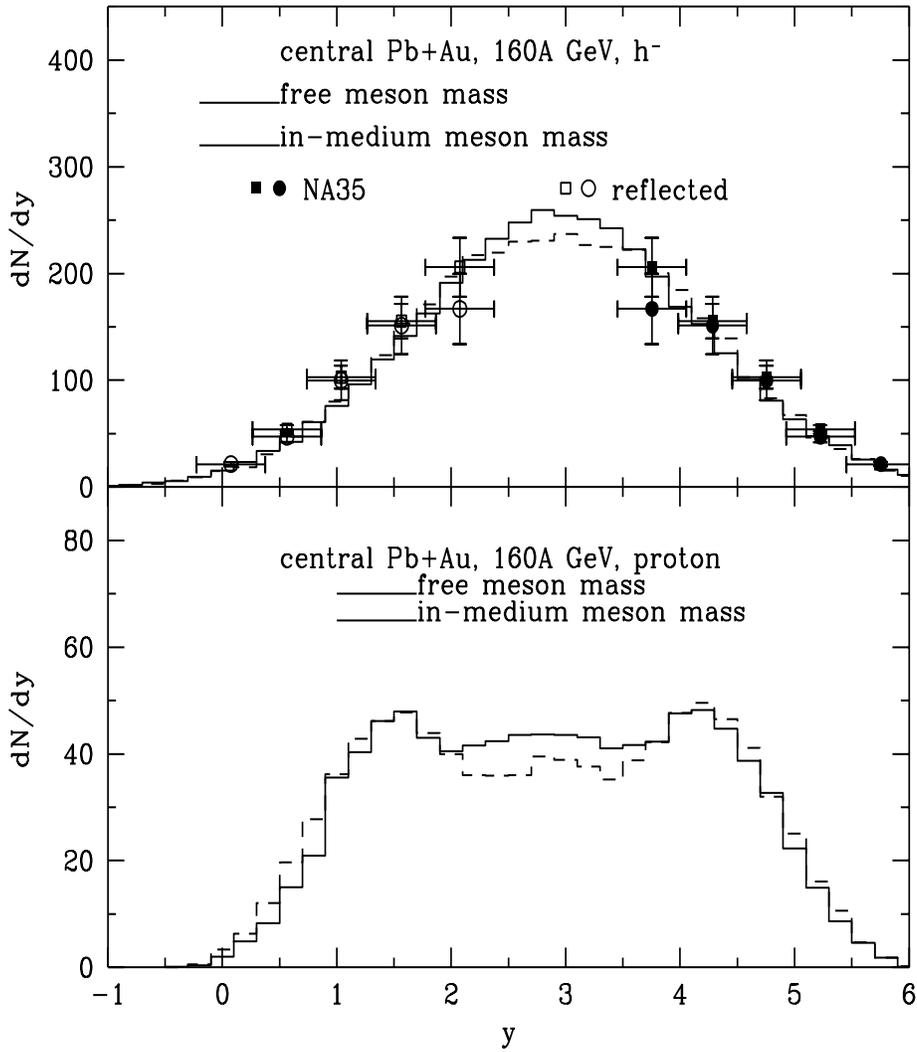,height=6in,width=5in}
\end{center}
\caption{The rapidity distributions of negatively-charged hadrons
and protons in central Pb+Au collisions at 160 GeV/nucleon.
Experimental data from the NA49 collaboration [61] are shown 
by circles.}
\end{figure}
 
\newpage

\begin{figure}
\begin{center}
\epsfig{file=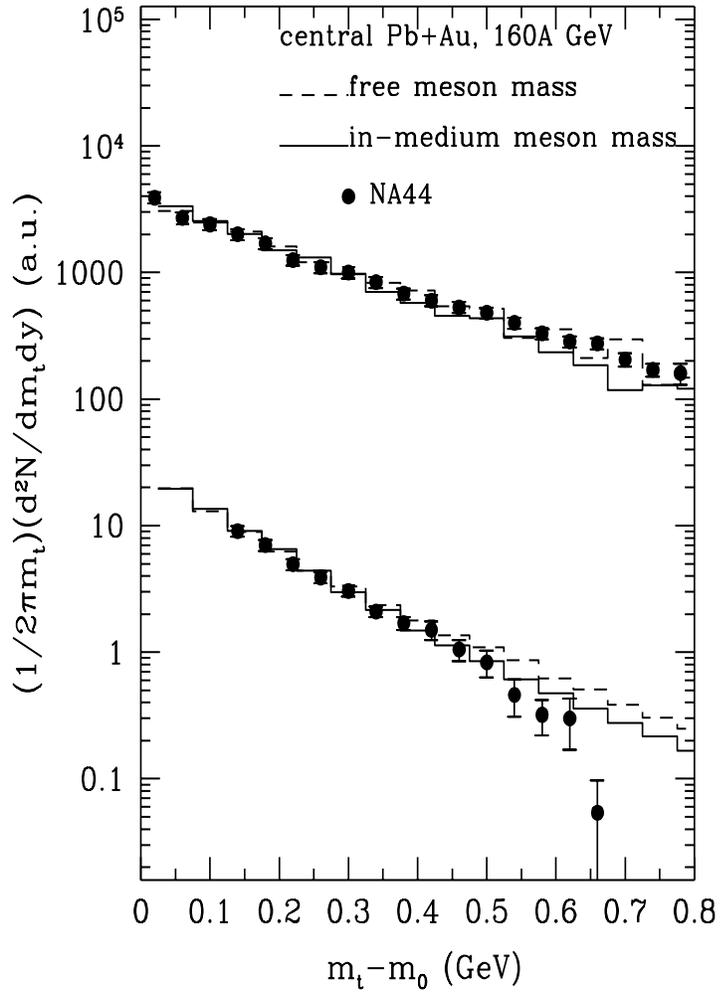,height=6in,width=5in}
\end{center}
\caption{The transverse mass spectra of protons and pions in 
the mid- to forward-rapidity region from central Pb+Au collisions at
160 GeV/nucleon. Experimental data from the NA44 collaboration 
[62] are shown by solid circles.}
\end{figure}

\newpage
 
\begin{figure}
\begin{center}
\epsfig{file=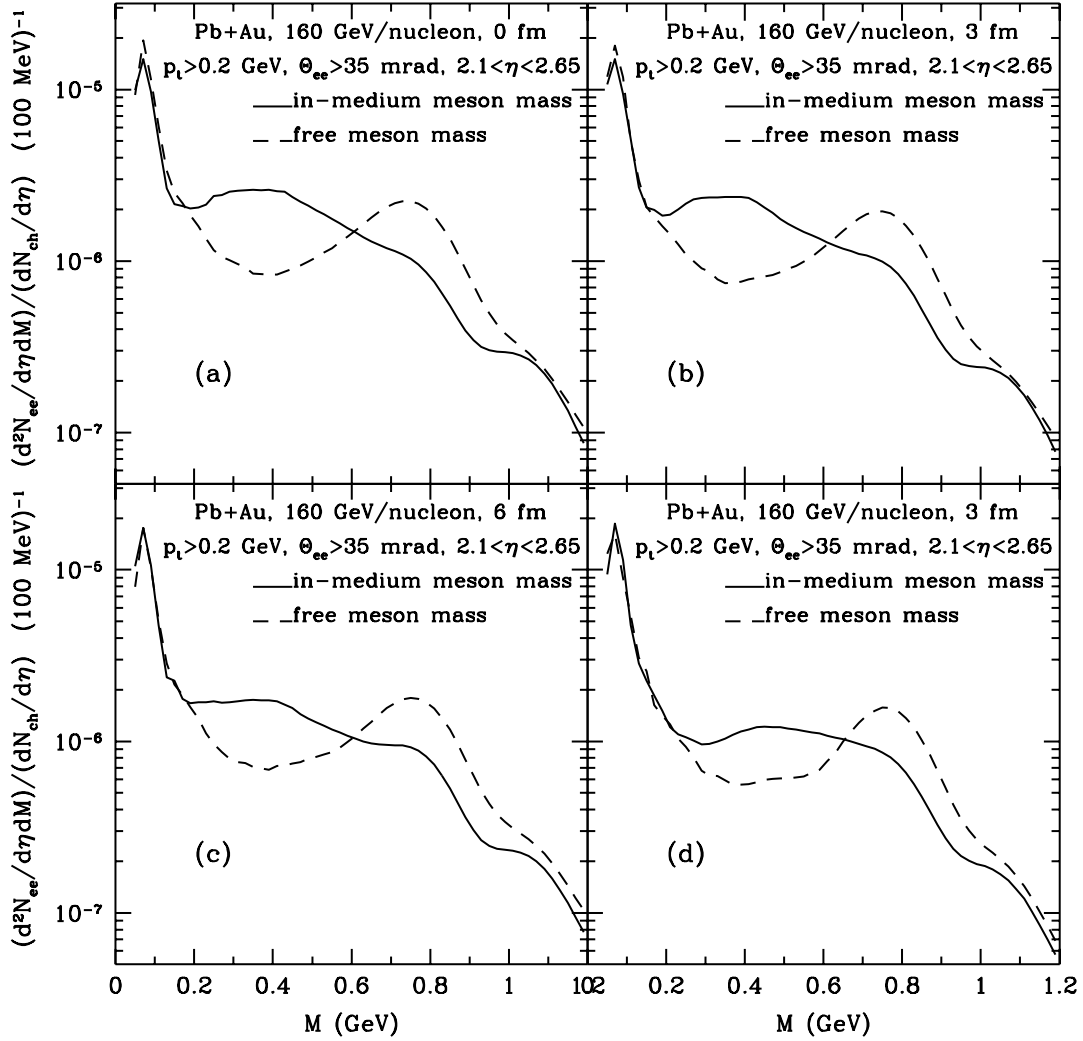,height=6in,width=6in}
\end{center}
\caption{Dilepton spectra from Pb+Au collisions with free 
(dashed curves) and in-medium (solid curves) meson masses
for the impact parameters (a) 0, (b) 3, (c) 6, and (d) 9 fm.
The CERES mass resolution and acceptance cuts 
for the S+Au collisions are included.}
\end{figure}

\newpage

\begin{figure}
\begin{center}
\epsfig{file=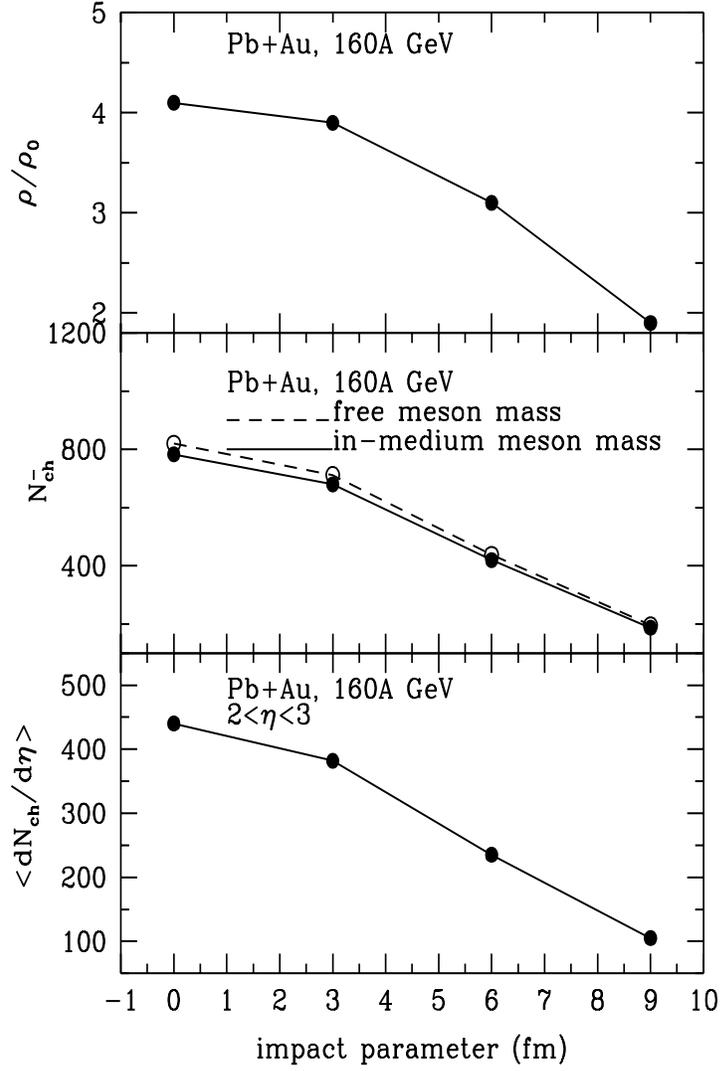,height=6in,width=5in}
\end{center}
\caption{Impact parameter dependence of the initial average
baryon density, the final negatively-charged particle multiplicity, and 
the charge particle pseudorapidity density at midrapidity for Pb+Au 
collisions at 160 GeV/nucleon.}
\end{figure}

\newpage
 
\begin{figure}
\begin{center}
\epsfig{file=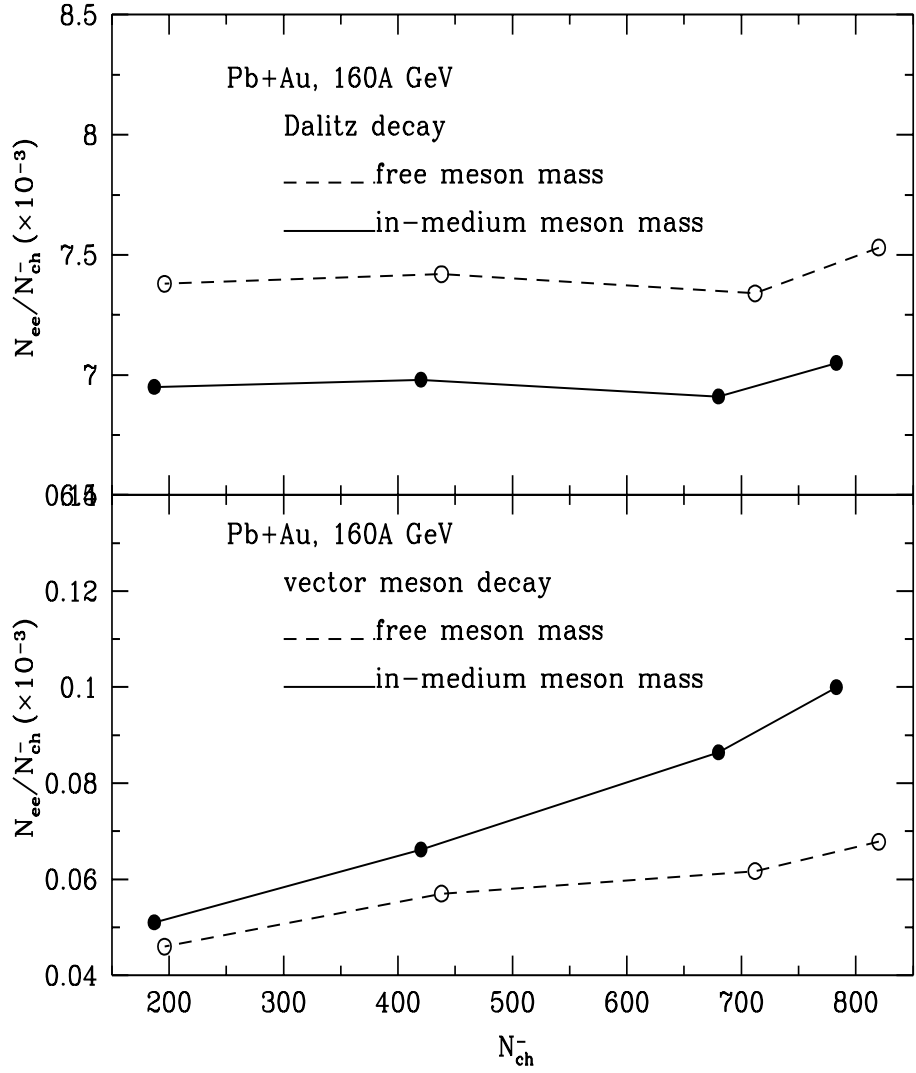,height=6in,width=5in}
\end{center}
\caption{The normalized total dilepton yield as a function of
the total negatively-charged hadron multiplicity in Pb+Au collisions 
at 160 GeV/nucleon.}
\end{figure}
 
\end{document}